\documentclass[12pt,oneside,notitlepage,abstracton,a4paper]{scrartcl}
\usepackage{epsfig,scrpage2,graphicx}

\lehead{EUDET-Report-2007-01}
\lohead{EUDET-Report-2007-01}

\titlehead{EUDET-Report-2007-01}
\subject{\includegraphics[bb=0 0 179 200,scale=0.7]{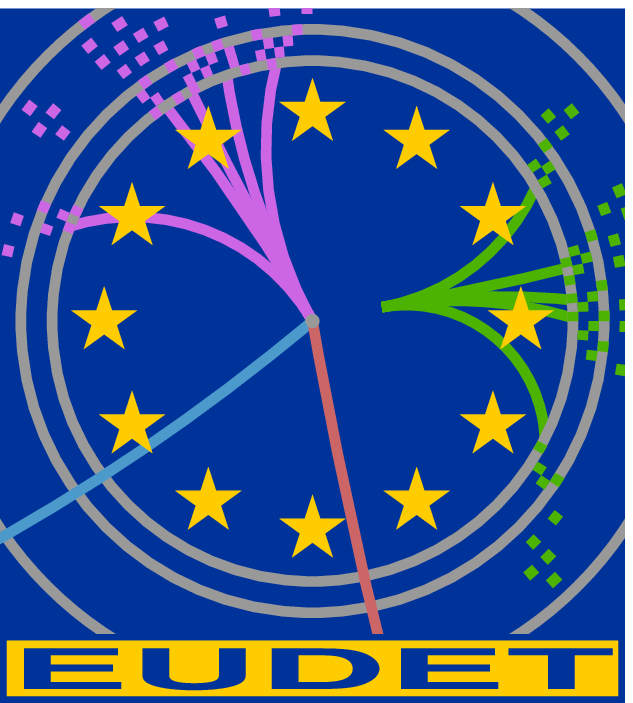}}

\def\eqa#1{\begin{eqnarray*} #1  \end{eqnarray*}}
\def\eqan#1{\begin{eqnarray} #1  \end{eqnarray}}

\def\bei{\begin{itemize}}
\def\eni{\end{itemize}}

\def\bec{\begin{center}}
\def\enc{\end{center}}

\def\eps{\varepsilon}

\def\fig#1{Figure~\ref{fig:#1}}
\def\eqref#1{(\ref{eq:#1})}

\def\sigdut{\sigma_{_{DUT}}}
\def\sighr{\sigma_{_{HR}}}
\def\deldut{\Delta_{_{DUT}}}
\newlength{\figheight}
\newlength{\diagheight}
\newlength{\diagsheight}
\newlength{\figwidth}
\newlength{\doublefigwidth}
\newlength{\quaterfigwidth}
\title{\Large EUDET Telescope Geometry and Resolution Studies} 
\author{\normalsize 
A.F.\.Zarnecki$^\ast$, 
P.Nie\.zurawski\thanks{Institute of Experimental Physics, 
                       Warsaw University, Warsaw, Poland}}
\date{\normalsize February 12, 2007}

\begin{document}

\maketitle

\begin{abstract}

\noindent
Construction of EUDET pixel telescope will 
significantly improve the test beam 
infrastructure for the ILC vertex detector studies.
The telescope, based on the Monolithic Active Pixel Sensors (MAPS),
will consist of up to six readout planes and will be 
initially installed at an electron beam line at DESY.
A dedicated study was performed to understand the position 
measurement in the telescope and optimize its performance 
by suggesting the best plane setup.
The approach based on the analytical track fitting method
allows for determination of the expected measurement uncertainty and for
comparison of different telescope setups without 
time consuming event simulation.
The proposed method also turns out to be little sensitive to
the possible telescope misalignment.

\end{abstract}

\newpage

\section{Introduction}

The goal of the EUDET project \cite{eudet} is the improvement 
of infrastructures to enable R\&D on detector 
technologies for ILC with large scale prototypes. 
The project is supported by the European Union in the 
Sixth Framework Programme structuring the European Research Area \cite{fp6}. 
One of the research activities, aiming at improving the test beam 
infrastructure for the vertex detector studies,
includes construction of a beam telescope based on 
Monolithic Active Pixel Sensors (MAPS) \cite{jra1}.
The telescope will consist of up to six sensor planes
and will be initially installed at a 6 GeV/c electron beam line at DESY. 
However, the device should be built in such a way that it
could later be used at other laboratories, for instance at high energy
hadron beam at CERN.
To meet all the requirements and provide a universal test environment 
for a wide variety of different sensor prototypes,
the telescope has to be designed very carefully.
One of the important issues is the choice of the telescope plane
configuration. 

The main aim of the presented study was to  
understand the position measurement in the telescope and
optimize its performance by suggesting the best plane setup.
A dedicated track fitting method was developed taking into
account multiple scattering.
Use of analytical approach allowed us to 
identify main factors determining the measurement error
and compare different telescope setups 
without time consuming MC simulations.

\section{Analysis method}

\subsection{Assumptions}

The approach presented in this paper is based on few simplifying 
assumptions. These assumptions seem to be realistic for the considered
telescope parameters. 

The most important assumption is that for the angle between 
the actual particle track and the nominal beam direction is always small. 
This is fulfilled if the incoming beam has a limited angular spread
and if the particle scattering angles in subsequent telescope layers are small. 
We also assume that the beam is perpendicular to the sensor planes, 
so that the track length between planes and the effective thickness 
of the material layer crossed by the particle are the same for all particles 
and given by telescope geometry.
Thicknesses of all material layers are very small compared 
to the distances between planes and we assume they can be neglected 
when calculating the particle scattering angle.

The distribution of the scattering angle is assumed to be Gaussian
and the expected width of the distribution can be estimated 
from the formula \cite{scattering}:
\eqan{ \Delta \Theta^{plane} & = & 
\frac{13.6 \; MeV}{\beta c p} \; z \; \sqrt{\frac{dx}{X_0}} \;
\left[ 1 \; + \; 0.038 \ln \left( \frac{dx}{X_0} \right) \right]
\label{eq:scat}
}
where $p$, $\beta c$ and $z$ are the momentum, velocity and charge of 
the incident particle, and $dx/X_0$ is thickness of the scattering medium 
in radiation lengths. For the telescope planes we use the radiation length 
of silicon, $X_0^{Si} = 9.36$ cm.
We also assume that the position measurement errors in all planes are Gaussian
and that there are no correlations between the horizontal and vertical position
measurements. 

With all the above assumptions the particle track reconstruction
in the telescope can be separated into two independent problems of 
track fitting in horizontal and vertical planes.

\subsection{Geometry description}

\begin{figure}[tbp]
  \begin{center}
     \epsfig{figure=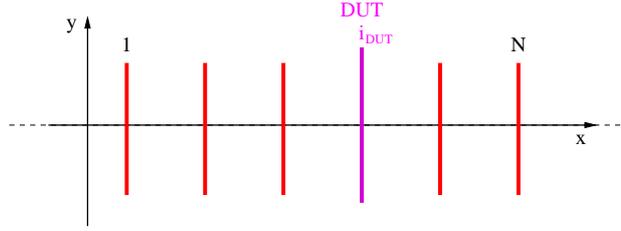,height=\diagheight,clip=}
  \end{center}
 \caption{
 Schematic view of the EUDET beam telescope setup. 
 }
 \label{fig:geometry2}  
 \end{figure}

The schematic view of the EUDET beam telescope is presented 
in \fig{geometry2}. 
We use the right-handed Cartesian coordinate system
with the $X$ axis pointing in the nominal beam direction and $Z$ axis 
pointing up.
In the presented analysis we use the following variables to 
describe the telescope geometry: 
\bei
\item $N$ - number of telescope planes;
all sensor planes as well as nonactive material layers and included,
position of the tested detector plane is denoted by $i_{_{DUT}}$,
\item $x_i$ - position of each plane along the beam direction ($i=1 \ldots N$),
\item $\Delta_i$ - thickness of each plane in radiation lengths,
\item $\Delta \theta_i$ - average scattering angle in each plane,
           calculated from formula \eqref{scat},
\item $\sigma_i$ - position measurement resolution (for sensor planes).
 \eni
%
%

In the following we will assume that the telescope consists only of
DUT and $N-1$ active sensor planes ($i\ne i_{_{DUT}}$) and there
is no additional nonactive material, which could affect the performance
of the telescope. 
We also assume that all detector layers are perfectly aligned.
The influence of the possible telescope misalignment and the effects
expected due to additional material layers in front of and behind DUT
(beam windows) will be discussed in the later part of this paper.

\subsection{Track fitting}

With the expected distances between telescope planes 
of the order of 10--100~mm and the scattering angles
of the order of 0.1~mrad (e.g. for 500$\mu m$ layer and 5 GeV electron beam),
the expected track displacement due to scattering can be of the
order of few micrometers.
It is comparable with the expected position resolution of
the telescope sensor layers ($\sim 2 \; \mu m$) and
can not be neglected.
The track fitting method described below takes it fully into account.
Similar approach was previously described in \cite{method}.


As mentioned above, track fitting can be considered separately for
horizontal and vertical plane (or any two perpendicular planes
parallel to the beam axis).
Therefor, in the description of the method we will limit ourselves 
to the two dimensional problem in the horizontal ($X-Y$) plane.

\begin{figure}[tbp]
  \begin{center}
     \epsfig{figure=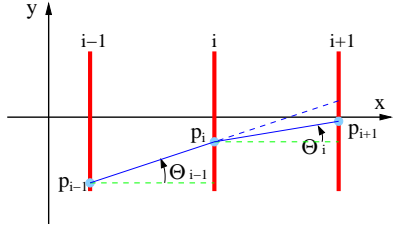,height=\diagheight,clip=}
  \end{center}
 \caption{
 Definition of angles used to calculate  particle multiple scattering 
in plane $i$. 
 }
 \label{fig:chi2} 
 \end{figure}

The aim of the fit is to determine particle positions in each telescope 
plane (including DUT), i.e. $N$ parameters ($p_i$, $i=1 \ldots N$), 
from $N-1$ positions measured in telescope planes
($y_i$, $i\ne i_{_{DUT}}$).
The problem can be solved, because we can use additional
constraints on the angles of multiple scattering.
Contribution of plane $i$ to $\chi^2$ of the fit can be written as:
\eqan{
 \Delta \chi^2_i & = & 
  \left. \left( \frac{y_i - p_i}{\sigma_i} \right)^2 \right|_{i\ne i_{DUT}}
    +  
  \left. \left( \frac{\Theta_i - \Theta_{i-1}}{\Delta \Theta_i} \right)^2 
    \right|_{i\ne 1,N}
\label{eq:chi2}
}
where $\Theta_i$ denotes the angle between the direction perpendicular 
to the telescope planes (nominal beam direction) and the particle track direction
between planes $i$ and $i+1$ (see \fig{chi2}).
The track angle can be calculated from particle positions in these planes:
\eqa{
 \Theta_i & = & \frac{p_{i+1} - p_i}{x_{i+1} - x_i} ~.
}
The first term in \eqref{chi2} is due to the uncertainty of the position
measurement and the second one reflects the expected distribution of the angle
of particle multiple scattering in plane $i$.
The measurement term drops out for DUT (and other nonactive telescope layers,
 if considered), whereas the scattering term is missing for the first and
the last plane, as the scattering angle can not be determined%
\footnote{
Constraint on the multiple scattering angle in the first plan can be included, 
if the angular spread of the incoming beam is known and can be approximated 
by the Gaussian distribution. This additional constrain influences the results
only if the angular spread of the incoming beam is comparable or smaller than 
the expected scattering angles in telescope planes. It was not considered 
in the presented analysis 
}.
%
%

After summing over all telescope layers we get general the following 
formula for $\chi^2$:
\eqa{
 \chi^2 & = & \sum_{i=1}^N \eps_i ( y_i - p_i )^2 
 \; + \; 
\sum_{i=2}^{N-1} \left( \frac{
(a_i + a_{i-1})p_i - a_{i-1} p_{i-1} - a_i p_{i+1}
}{\Delta \Theta_i} \right)^2
}
where the coefficients $\eps_i$ and $a_i$ are defined as
\eqa{
\eps_i & = & \left\{ \begin{array}{ll}
\frac{1}{\sigma^2_i} & \textrm{for ~} i\ne i_{_{DUT}} \textrm{~~(sensor planes)} \\
0 &  \textrm{for ~} i = i_{_{DUT}} 
\end{array} \right.   \\
a_i & = & \frac{1}{x_{i+1} - x_i}
}

Fitting a track implies finding most probable values of particle positions 
in telescope planes, i.e. finding minimum of $\chi^2$.
It is equivalent to solving the set of $N$ equations:
\eqa{\frac{\partial \chi^2}{\partial p_i} & = & 0,  ~~i=1 \ldots N. }
The advantage of the presented approach is that the derivatives of $\chi^2$
can be calculated analytically. First derivatives are linear functions of 
particle positions $p_i$ and the problem can be further reduced to solving 
a matrix equation:
\eqa{ \sum_j {\cal A}_{ij}\; p_j & = & \eps_i \; y_i }
where:
\eqa{ {\cal A}_{ij} & = & 
\frac{1}{2}\; \frac{\partial^2 \chi^2}{\partial p_i \; \partial p_j}}

The general formula for ${\cal A}$ is quite complicated,
but the matrix can be calculated analytically.
Moreover, all elements of ${\cal A}$ depend only on the assumed telescope 
geometry and do not depend on the measured particle positions $y_i$.
Therefor the matrix ${\cal A}$ needs to be calculated only once.
To solve the equation (i.e. fit the track)
we only need to find the inverse matrix:
\eqa{ {\cal S} & = & {\cal A}^{-1} }
which can then be used to fit tracks for all collected events.
Reconstructed particle position in plane $i$ is given by
a linear combination of measured positions $y_j$ in all active layers:
\eqan{ p_i & = & \sum_j  {\cal S}_{ij} \; \eps_j \; y_j
             \label{eq:fit} }
Position measurement at DUT, although formally included in the general formula, 
does not contribute to the results as $\eps_{i_{DUT}} = 0$ 
(same is true for other nonactive telescope layers, if considered).

The diagonal elements of ${\cal S}$ correspond to the expected
precision of the particle position reconstruction.
The error on the particle position at plane $i$  is given by:
\eqa{ \tilde{\sigma}_{i} & = & \sqrt{{\cal S}_{i\; i}  } }
Again, it turns out that the uncertainties of reconstructed particle
positions depend only on the assumed telescope geometry and do not 
depend on the measured particle positions. 
This feature of the proposed analysis method is of great importance, 
as it allows us to compare the precision of position determination at DUT,
$\sigdut \equiv  \tilde{\sigma}_{i_{DUT}}$, for different
telescope geometries and beam energies, without time consuming 
event generation based on Monte Carlo methods.


\section{Simulation results}
\label{sec:sim}

\subsection{Telescope configuration}

\begin{figure}[tbp]
  \begin{center}
     \epsfig{figure=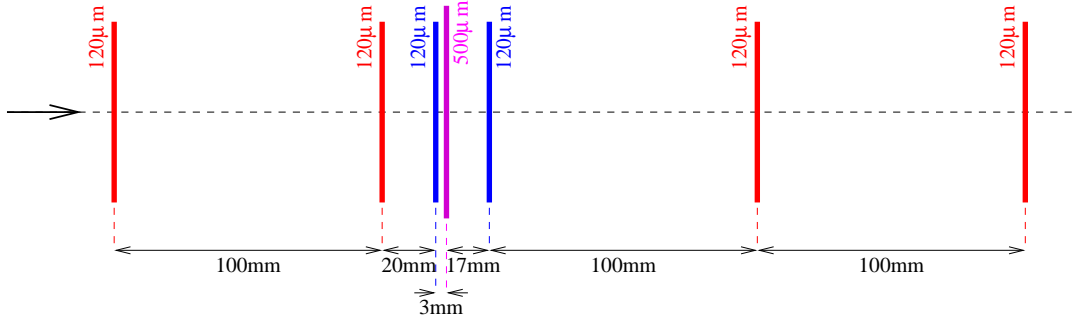,width=0.9\textwidth,clip=}
  \end{center}
 \caption{
 The telescope configuration as used in the simulation studies.
 }
 \label{fig:conf_wn-ww_2} 
 \end{figure}

\begin{figure}[tbp]
  \begin{center}
     \epsfig{figure=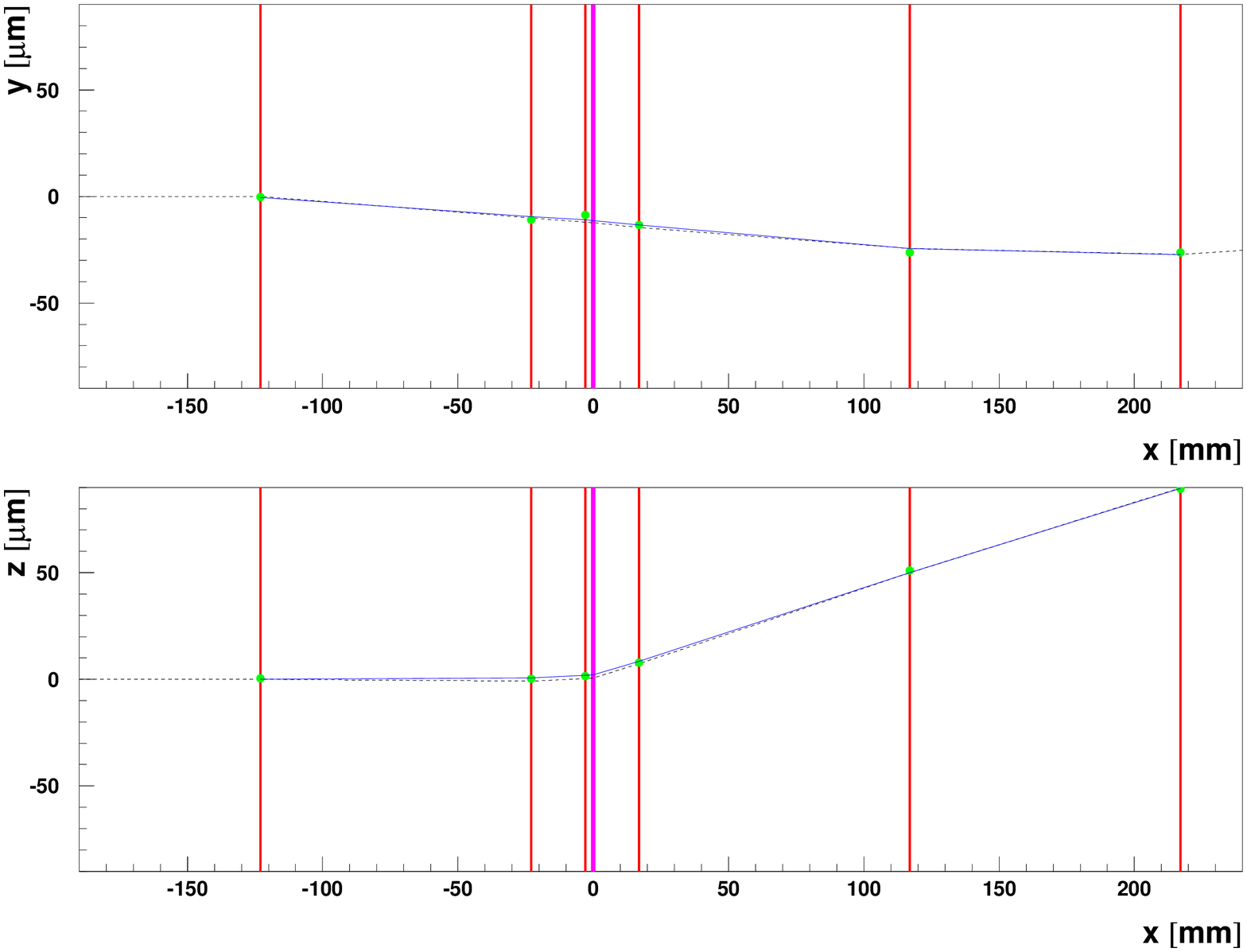,height=\figheight,clip=}
  \end{center}
 \caption{
 Simulation results in horizontal (upper plot) and vertical (lower plot)
 plane, for 6~GeV electron passing through the telescope.
 Black dashed line corresponds to the simulated particle trajectory,
  green dots represent the simulated position measurements, blue solid
  line is the output of the described reconstruction method. 
 }
 \label{fig:geant_event_1} 
 \end{figure}

To verify the validity of different assumptions used in the proposed 
track fitting method, a dedicated simulation study based on {\sc Geant~4}
has been performed.
The telescope configuration used in the simulation is presented in
\fig{conf_wn-ww_2}.
%
%
Telescope consists of 7 layers: 6 sensor layers with thickness of 120$\mu m$
and a DUT, 500 $\mu m$ thick, placed as a middle layer (indicated in magenta).
The first two and the last two sensor planes (indicated in red) are assumed 
to have the position resolution of $2 \mu m$ (standard sensor planes),
whereas for the sensor planes adjacent to DUT (indicated in blue)
resolution of $1 \mu m$ is assumed (high-resolution planes).
Distances between planes, as indicated in the plot, correspond to 
the configurations which turned out to result in best particle position 
determination at DUT.
Detector response is simulated by a Gaussian smearing of the particle position 
with the assumed detector resolution.
True particle position at given layer is taken as the mean value of the 
positions at the entry and exit points, as obtained from simulation.
Simulation was performed for two incident beam energies: 
electron beam of 6~GeV and pion beam of 100~GeV.
A sample event from the 6~GeV electron beam simulation is presented 
in \fig{geant_event_1}.
%
%

\subsection{Reconstruction results}

\begin{figure}[tbp]
  \begin{center}
     \epsfig{figure=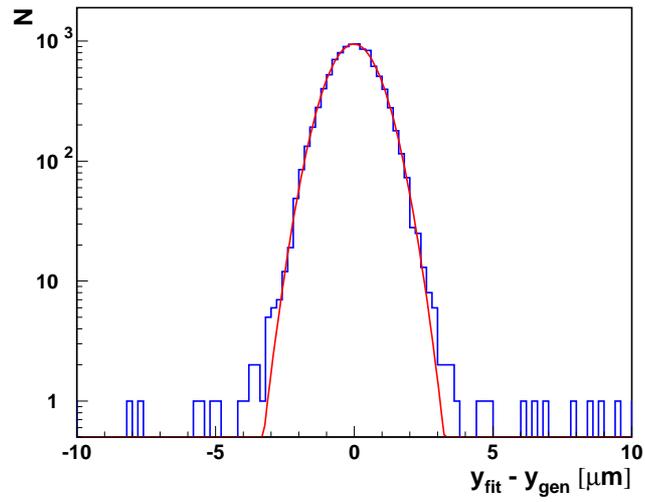,height=\figheight,clip=}
  \end{center}
 \caption{
 Reconstruction error distribution for the particle position at DUT, for
6 GeV electron beam. Simulation results (blue histogram) are compared with 
the fitted Gaussian distribution (red line).
 }
 \label{fig:shift} 
 \end{figure}

Also shown in \fig{geant_event_1} is the result of particle track 
reconstruction with the described method (solid blue line).
Reconstructed track follows very closely the "true" particle trajectory 
(indicated by dashed line) properly describing particle deflection due to 
multiple scattering both in DUT and in the sensor layers.
The difference between the reconstructed and generated particle position at DUT
is presented in \fig{shift}.
As expected, distribution of the position reconstruction error has a Gaussian
shape, as indicated by the fitted  Gaussian distribution.
Only very few events populate non-Gaussian tails, which are probably due
to the events with large multiple scattering, not described by the approximate
formula \eqref{scat}.
%
%

\begin{figure}[tbp]
  \begin{center}
     \epsfig{figure=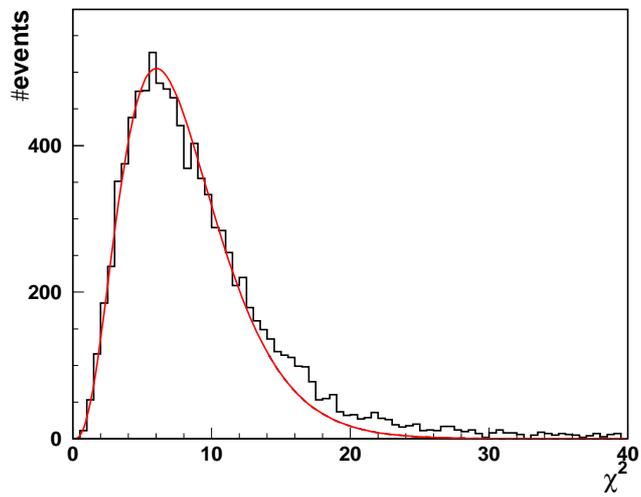,height=\figheight,clip=}
  \end{center}
 \caption{
 Distribution of the total $\chi^2$ of the track fit,
 for 6~GeV electron beam, compared with
the expected $\chi^2$ distribution for 8 degrees of freedom.
}
 \label{fig:chi2fit} 
 \end{figure}

The quality of the track reconstruction is also illustrated in \fig{chi2fit}.
Distribution of the total $\chi^2$ of the track, calculated as a sum 
of $\chi^2$ from fits in the horizontal and vertical plane, is compared
with the expected $\chi^2$ distribution.
Results of the simulation are very well described by the distribution 
expected for 8 degrees of freedom.
This is consistent with our expectations, as we fit 14 parameters
(particle position in $Y$ and $Z$, in 7 telescope planes) using 
12 measurements (from 6 sensor planes) and 10 constraints on 
multiple scattering (5 inner telescope planes).
In the general case, number of degrees of freedom is given by
\eqa{ N_{df} & = & 2 \cdot N_s \; - \; 4, }
where $N_s$ is the number of active sensor layers.
%
%

\begin{figure}[tbp]
  \begin{center}
     \epsfig{figure=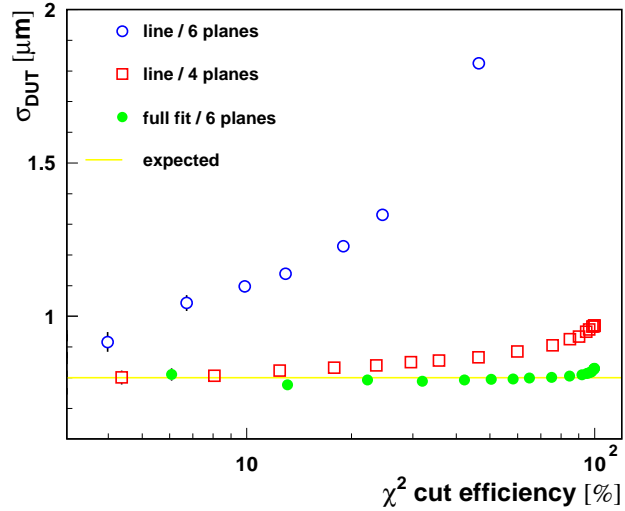,height=\figheight,clip=}
  \end{center}
 \caption{
Obtained position resolution at DUT, $\sigdut$,
as a function of the fraction of events accepted after $\chi^2$,
Results from the described method taking into account multiple
scattering (green points) are compared with position resolution obtained from
the straight line fit to 6 sensor planes and to 4 inner sensor planes.
 }
 \label{fig:chi2cut} 
 \end{figure}

Precision of particle position determination at the DUT plane,
as obtained from the Gaussian distribution fit to all simulated
events (see \fig{shift}) is 0.84$\mu m$.
However, for few percent of events we observe $\chi^2$ 
values larger than expected (see  \fig{chi2fit}).
If we reject these events, position reconstruction uncertainty 
of about 0.81$\mu m$ is obtained, in very good agreement with 
0.80$\mu m$ expected from the analytical calculation.
The dependence of the obtained position resolution at DUT, $\sigdut$,
on the fraction of events accepted after $\chi^2$ cut, is presented in
\fig{chi2cut}.
After rejecting of about 10\% of events with highest $\chi^2$ values, 
telescope position resolution remains  flat and agrees very well
with the expected precision (indicated by yellow line).
For comparison, position resolution at DUT, as obtained from
the straight line fit to 6 sensor planes and to 4 inner sensor planes
is also shown in \fig{chi2cut}.
Line fit to 4 planes results in much better position determination than 
the fit to 6 planes.
This is because the measurements in the first and the last plane are
least correlated with the particle position at DUT due to
effects of multiple scattering.
Reconstruction method presented in this paper takes these effect
into account and allows us to obtain optimum position measurement for
90\% of events, whereas for the line fit the corresponding precision could
only be reached for about 10\% of events with smallest multiple scattering.
%
%

\subsection{Comparison with analytical error estimates}
\label{sec:geantcomp}

\begin{figure}[tbp]
  \begin{center}
     \epsfig{figure=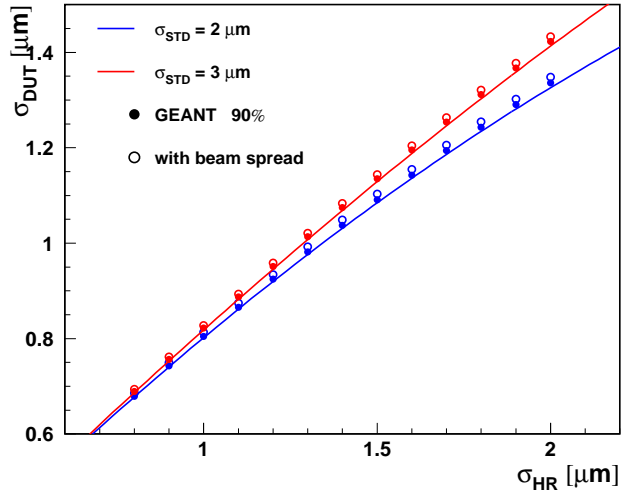,height=\figheight,clip=}
  \end{center}
 \caption{
 Precision of position determination at DUT, for 6~GeV electron beam, 
 as a function of the  position resolution in the high-resolution planes, 
$\sighr$, for standard planes resolution of $2 \mu m$ (blue) 
and $3 \mu m$ (red). Results of the {\sc Geant 4} simulation (points) are
compared with expectations from the analytical calculations (lines).
 }
 \label{fig:geantres} 
 \end{figure}

Results presented so far were based on the simulation assuming
position measurement resolution of $1 \mu m$ for two sensor planes 
adjacent to DUT (high-resolution planes) and $2 \mu m$ for the remaining
standard sensor planes.
To test the validity of analytical approach to position error determination, 
simulation of detector response was repeated for different values of
the assumed sensor plane position resolutions.
Shown in \fig{geantres} is the precision of position determination at DUT, 
$\sigdut$, expected for 6~GeV electron beam, as a function of
the  position resolution in the high-resolution planes, $\sighr$.
Results of the {\sc Geant 4} simulation, for 90\% of events, i.e. after rejecting
10\% of events with the highest $\chi^2$ values (points) are compared with the
resolution expected from the analytical calculations (lines).
Standard planes resolution of $2 \mu m$ and $3 \mu m$ is considered.
Simulation results agree very well with the model expectations if 
no angular spread of the incident electron beam is included simulation (filled points).
When the angular spread of 0.1$\mu m$ is assumed (open points),
precision of position determination obtained from the simulation 
worsens by 1--2\%. 
This indicates the level of systematic uncertainty of the analytical error
estimate due to the simplifying assumptions used in the described method.
%
%

\begin{figure}[tbp]
  \begin{center}
     \epsfig{figure=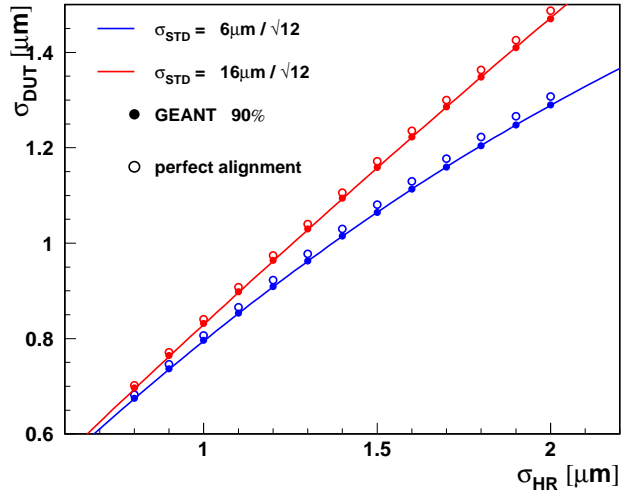,height=\figheight,clip=}
  \end{center}
 \caption{
 Precision of position determination at DUT, for 6~GeV electron beam, 
 as a function of the  position resolution in the high-resolution planes, 
$\sighr$, for standard planes with binary readout and pitch of 
$6 \mu m$ (blue) and $16 \mu m$ (red). Results of the {\sc Geant 4} simulation 
(points) are compared with expectations from the analytical calculations (lines).
 }
 \label{fig:geantbin2} 
 \end{figure}

Although the method is based on the assumption of the Gaussian position
resolution in each plane, this assumption is not crucial for the reliable
error estimates.
This is demonstrated in figure \fig{geantbin2}.
As before,  precision expected from analytical calculation
is compared with results of {\sc Geant 4}  simulation.
However, the position measurement error in the standard planes 
is modeled by the uniform distribution, corresponding to 
the binary readout option considered for these planes.
The error used in analytical calculations is set to
$\frac{1}{\sqrt{12}}$ of the distribution width, which
is given by the sensor pitch of $6 \mu m$ or $16 \mu m$.
%
%

\begin{figure}[tbp]
  \begin{center}
     \epsfig{figure=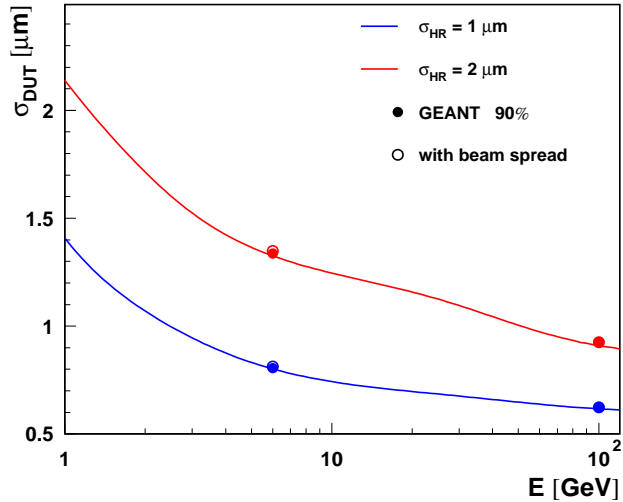,height=\figheight,clip=}
  \end{center}
 \caption{
 Precision of position determination at DUT, for  position resolution 
in the high-resolution planes of $1 \mu m$ (blue) and $2 \mu m$ (red),
as a function of the incident beam energy.
Results of the {\sc Geant 4} simulation (points) are
compared with expectations from the analytical calculations (lines).
 }
 \label{fig:eneres} 
 \end{figure}

To verify the description of the energy dependence, the {\sc Geant 4} 
simulation was also performed for 100~GeV pion beam.
Dependence of the expected position uncertainty at DUT
on the incident beam energy is shown in \fig{eneres}.
The agreement of analytical error estimates with the uncertainty calculated
from the  {\sc Geant 4} simulation is very good.
%
%


\section{Telescope configuration studies}

As demonstrated in the previous section, presented method 
gives reliable results for the precision of position measurement
at DUT.
The position error depends only on the assumed telescope parameters
and can be calculated without any event simulation.
This allows us to make a quantitative comparison between different
telescope configurations.
We tried to optimize the telescope setup, by finding telescope
geometry which results with the best position determination at DUT.
For given number and resolutions of sensor planes,
order and distances between planes were varied to find minimum 
of $\sigdut$.
If no constraints are put on plane separation,
smallest error is always obtained when all distances go to zero.
Therefor, to obtain realistic results, one has to take into account
constraints which result from the mechanical structure of the telescope.

\subsection{Realistic telescope geometry}

\begin{figure}[tbp]
  \begin{center}
     \epsfig{figure=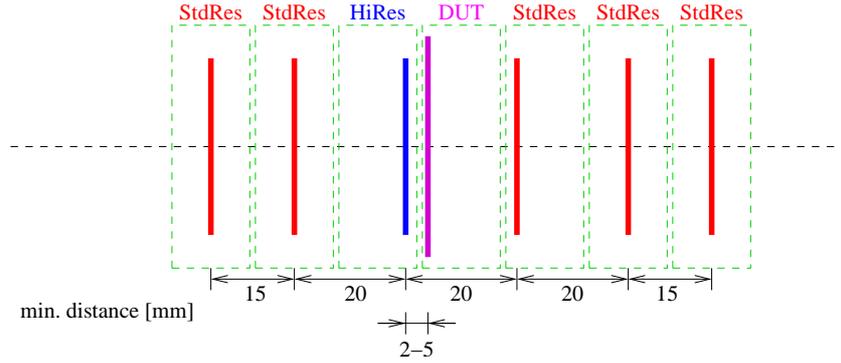,width=0.7\textwidth,clip=}
  \end{center}
 \caption{
Schematic view of the realistic configuration for the telescope
consisting of DUT (magenta), one high-resolution sensor plane (blue)
and 5 standard sensor layers (red). Minimum distances between sensor planes 
are indicated. 
 }
 \label{fig:new7_1hr} 
 \end{figure}

Distances between planes are mainly restricted by the mechanical
structure of their support frames. 
For standard sensor layers the minimum distance between two sensors
is expected to be of the order of 15~mm. 
It is assumed that the telescope should comprise at least one 
high-resolution plane.
This plane could be placed as close as 2--5~mm to DUT.
We assume that it will be places in front of DUT%
\footnote{As we do not consider beam energy losses in the telescope planes,
same results are obtained if the order of all planes is reversed.}.
Minimum distance between two planes adjacent to DUT is about 20~mm.
\fig{new7_1hr} shown the schematic view of the telescope with
minimum distances between sensor planes indicated.
As for the maximum distance, we assume that two subsequent planes should 
not be separated by more than 100~mm.
%
%

The choice of optimum telescope configuration was studied
for telescope consisting of 4 or 6 sensor planes (and DUT).
Moreover, possibility of having one or two high-resolution 
sensor planes was considered.
General conclusion from all studies is that DUT should always be placed
between two sensor planes, and the distance between these two planes
should be as short as possible. 
In most cases, to get the best position measurement
separations between other sensor planes should either be equal to the minimum 
or to the maximum allowed distance between them, depending on the energy, sensor
and DUT parameters.
High-resolution plane is always the one closest to DUT (we assume that 
These observation led to the following scheme for the labeling of telescope
configurations: 
\bei
\item DUT and two sensor planes close to it are always denoted by {\bf --},
\item any other sensor plane in front of or behind these 3 planes is
      denoted by {\bf N}, if placed at the minimum allowed distance 
       (narrow configuration)
\item or by {\bf W}, if placed at the maximum distance 
       (wide configuration).
\eni

All results presented in this section were obtained assuming the
thickness of telescope sensors of 120 $\mu m$, both for standard
and for high-resolution planes.
If not stated otherwise, the resolution of standard sensor planes
is assumed to be  2 $\mu m$.
For high-resolution sensors the default position determinations
uncertainty considered is 1 $\mu m$.

\subsection{Results for 1+3 configuration}

\begin{figure}[tbp]
  \begin{center}
     \epsfig{figure=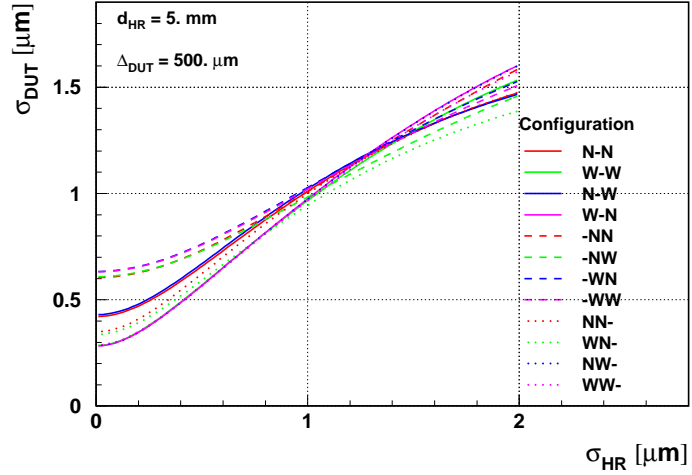,height=\figheight,clip=}
  \end{center}
 \caption{
Expected precision of position determination
at DUT, $\sigdut$, as a function of the position resolution in 
high-resolution plane, $\sighr$, for different telescope configurations,
for 6~GeV electron beam.
 }
 \label{fig:res5_1hr_sig} 
 \end{figure}

\begin{figure}[tbp]
\begin{center}
\begin{minipage}[t]{0.7\textwidth}
{\bf W--W}\\[-1cm]
\hspace*{2cm}
 \epsfig{figure=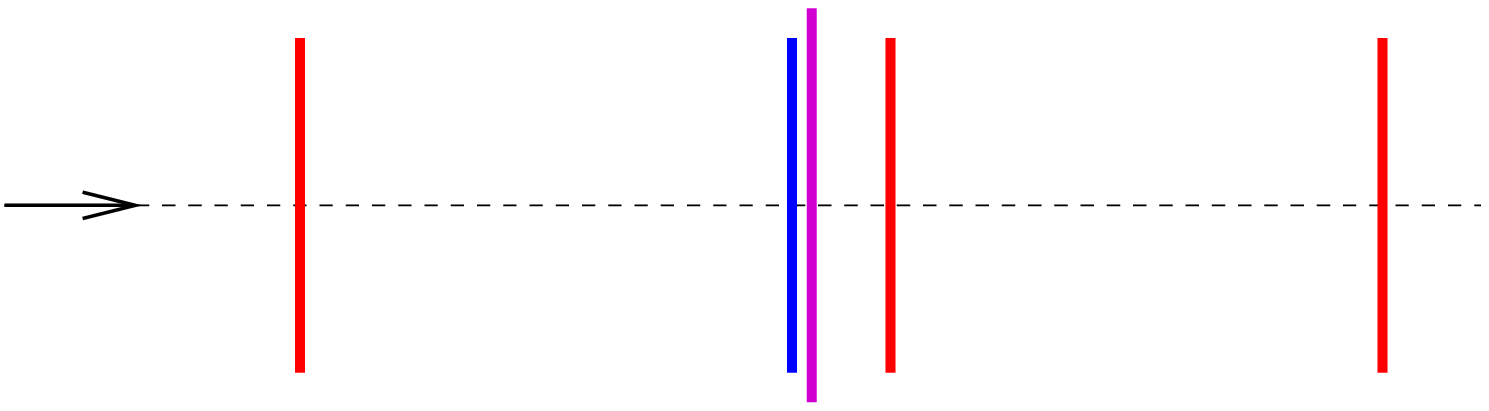,height=\diagsheight,clip=}\\[1cm]
{\bf WN--}\\[-1cm]
\hspace*{2cm}
 \epsfig{figure=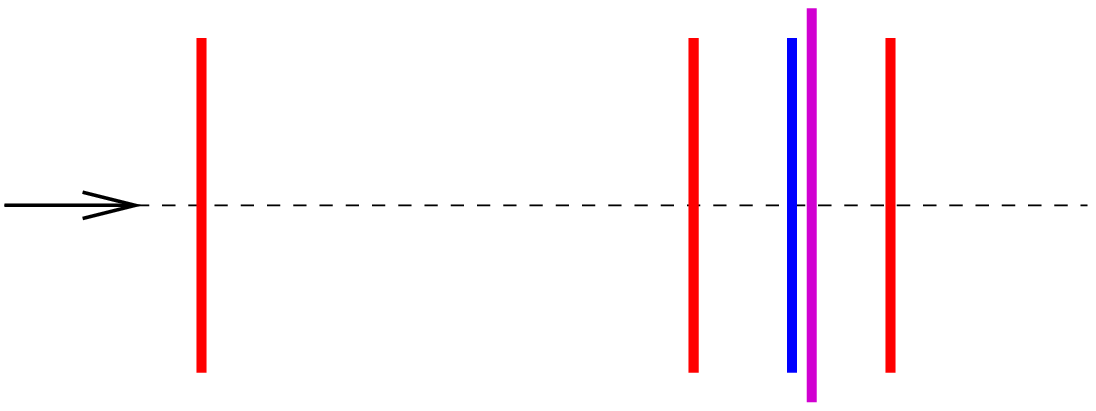,height=\diagsheight,clip=}\\[1cm]
{\bf N--N}\\[-1cm]
\hspace*{2cm}
 \epsfig{figure=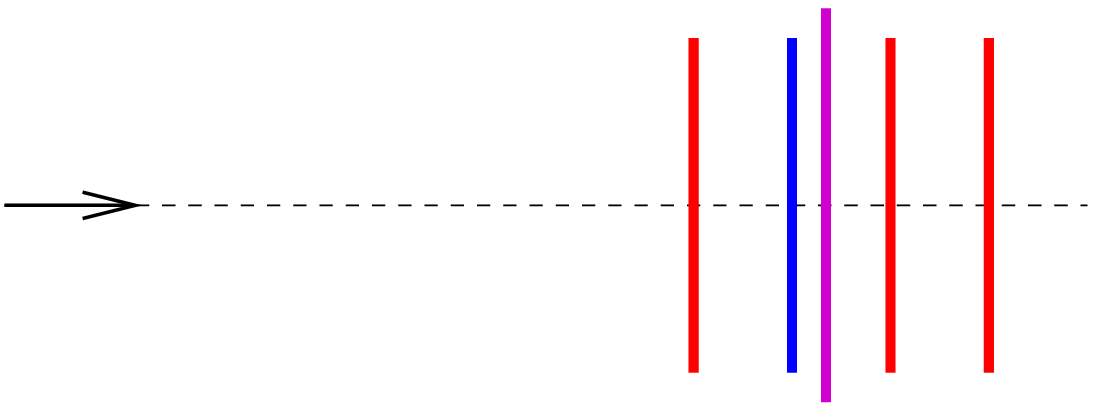,height=\diagsheight,clip=}
\end{minipage}
\end{center}
 \caption{
Configurations  resulting in the best position measurement
at DUT (in different ranges of parameters), 
for telescope consisting one high-resolution and three standard
sensor planes. High-resolution plane is indicated in blue, standard planes 
in red and the DUT in magenta.
 }
 \label{fig:conf_51} 
 \end{figure}

This section describes the results obtained for the telescope consisting
of one  high-resolution plane (placed in front of DUT) and three standard 
sensor planes (1+3 configuration).
\fig{res5_1hr_sig} shows the expected precision of position determination
at DUT, $\sigdut$, as a function of the position resolution in 
high-resolution plane, $\sighr$, for different telescope geometries,
for 6~GeV electron beam.
Minimum distance between DUT and high-resolution plane is set to be 5~mm
and the DUT thickness assumed is $\deldut = 500\mu m$.
For $\sighr \ge 1 \mu m$ the best measurement is obtained in the {\bf WN--}
configuration, i.e. when two standard and a high-resolution plane are 
placed in front of DUT and only one plane behind it.
The configuration scheme is shown in \fig{conf_51} (middle plot). 
%
%
%
%

\begin{figure}[tbp]
  \begin{center}
     \epsfig{figure=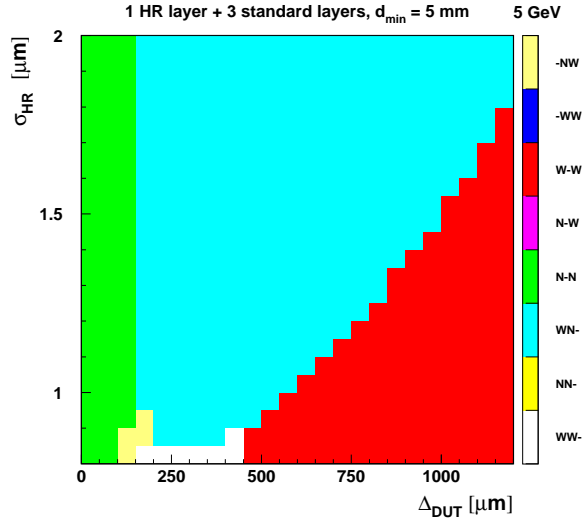,height=\figheight,clip=}
  \end{center}
 \caption{
Telescope configuration resulting in the is best position measurement
at DUT,  as a function of the assumed DUT thickness, $\deldut$ and the high-resolution plane measurement precision, $\sighr$, for electron beam
energy of 5~GeV.  Configurations with one high-resolution and three standard
sensor planes are considered. 
 }
 \label{fig:conf5_5_2_5} 
 \end{figure}

However, this configuration choice is optimal only for the limited range 
of $\sighr$ and  $\deldut$, and of the beam energy.
This is shown in \fig{conf5_5_2_5}, where the best choice of
telescope configuration is shown as a function of the assumed DUT 
thickness, $\deldut$ and the high-resolution plane measurement precision,
$\sighr$, for electron beam energy of 5~GeV. 
For very thin DUT, $\deldut < 150 \mu m$, when the effects of multiple 
scattering are small, the best measurement precision
can be obtained with {\bf N--N} configuration. On the other hand,
for very thick DUT, when multiple scattering gives 
dominant contribution to the position uncertainty, and for
small $\sighr$ values {\bf W--W} layout turns out to be 
a preferred choice (for configuration schemes refer to \fig{conf_51}).
%
%

\begin{figure}[p]
  \begin{center}
     $\Delta_{_{DUT}} = 800 \mu$m \\[-0.5cm]
     \epsfig{figure=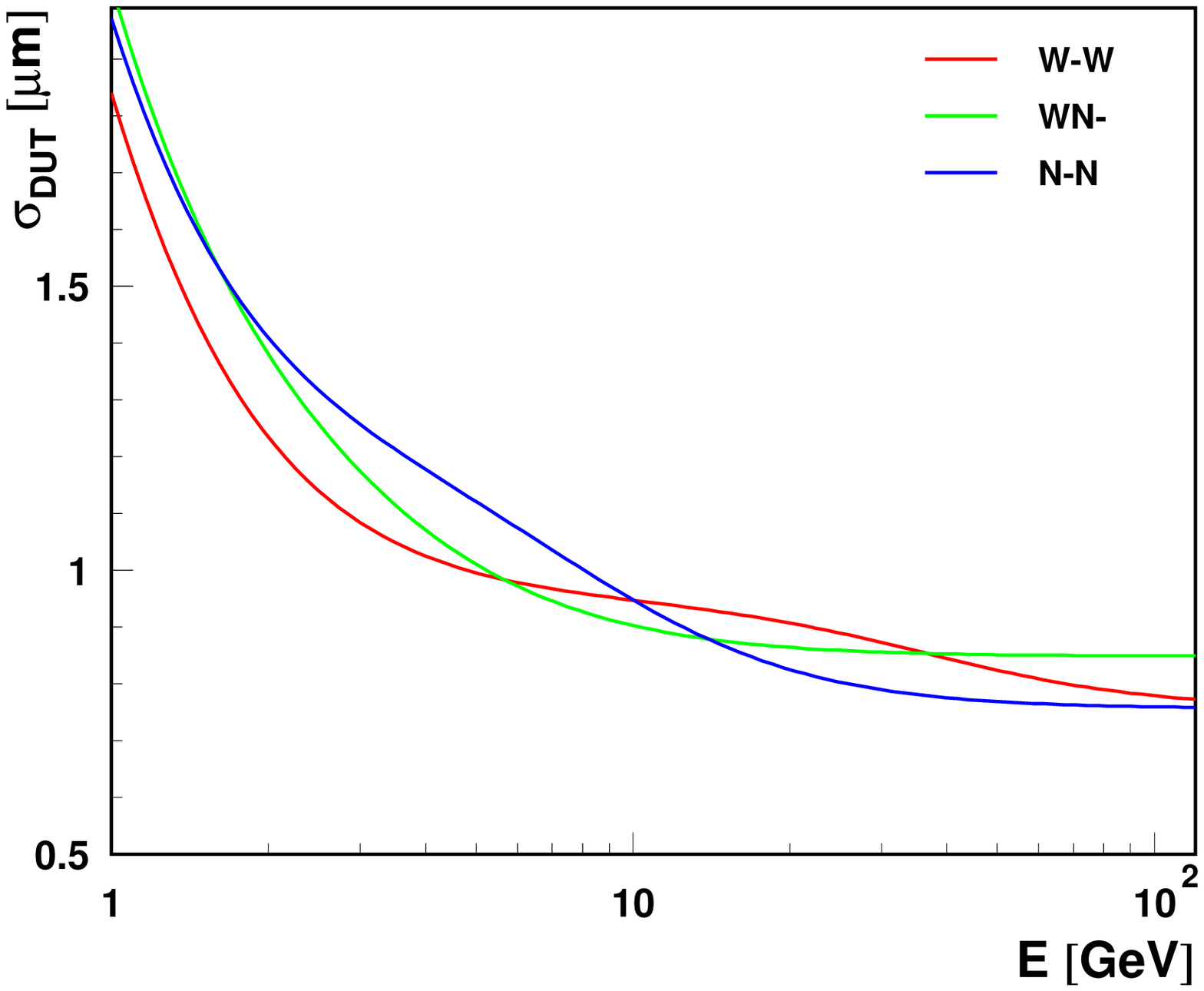,height=\figheight,clip=}\\
     $\Delta_{_{DUT}} = 300 \mu$m \\[-0.5cm]
     \epsfig{figure=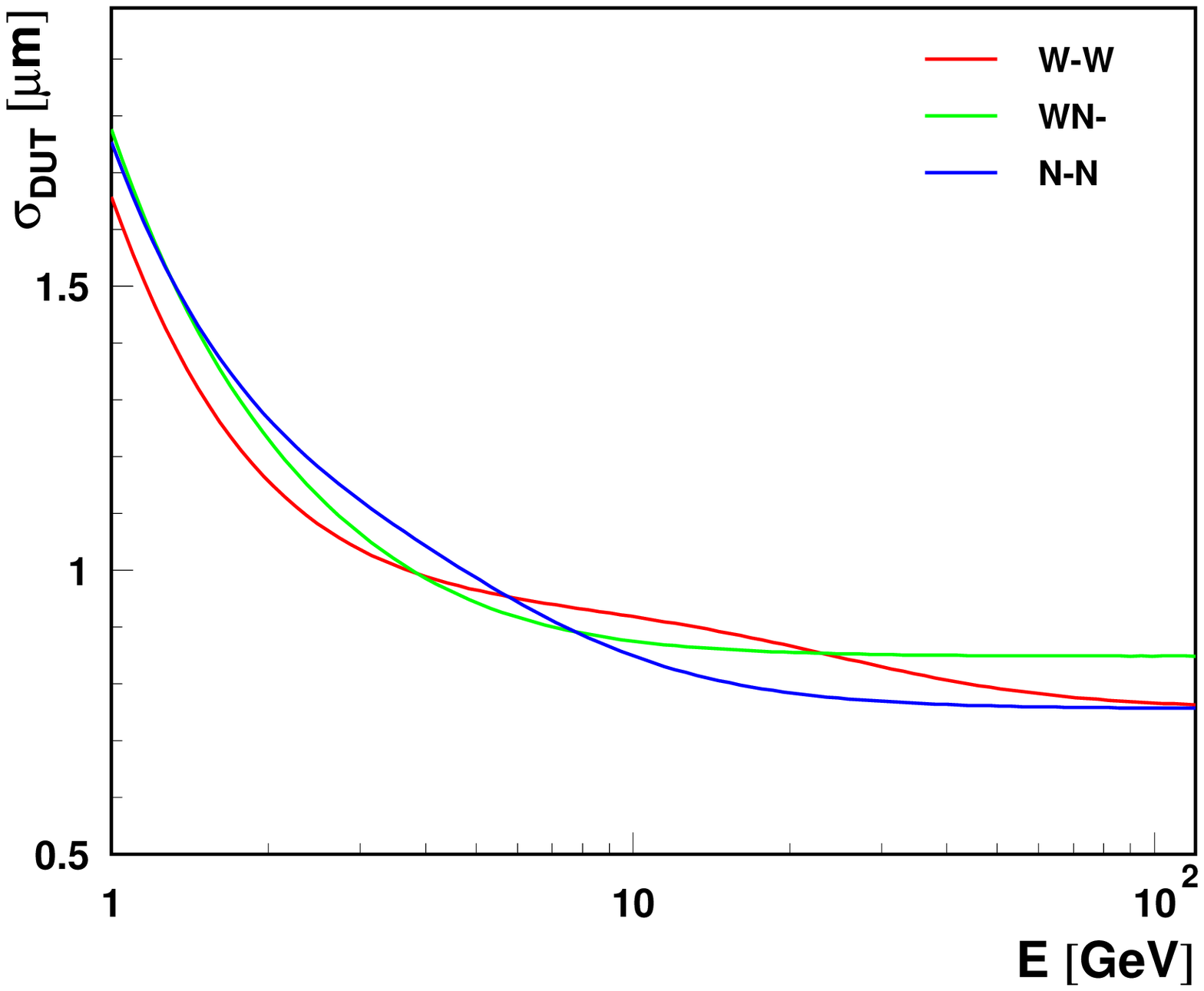,height=\figheight,clip=}
  \end{center}
 \caption{
Expected precision of position determination at DUT, $\sigdut$, 
as a function of the beam energy, 
for the assumed DUT thicknesses of 800$\mu m $ (upper plot) 
and 300$\mu m$ (lower plot) and the distance between 
the high-resolution plane and DUT of 5~mm.
Configurations with one high-resolution and three standard
sensor planes are considered. 
 }
 \label{fig:enecomp51_5} 
 \end{figure}

The position uncertainty due to multiple scattering depends on 
the particle energy.
The expected precision of position determination
at DUT, $\sigdut$, as a function of the beam energy, for the three
selected telescope configurations is shown in \fig{enecomp51_5},
for the assumed DUT thicknesses of 800 and 300$\mu m$,
and the distance between the high-resolution plane and DUT of 5~mm.
{\bf W--W} layout gives the best precision measurement at lowest
energies, when the effects of multiple scattering are dominant.
Large distances between sensor planes allow for a better estimate 
of the particle track direction before and after DUT, resulting in better
extrapolation of the position measured in the high-resolution plane
and in the plane behind DUT to the DUT surface.
At high energies influence of multiple scattering is small and
the position error at DUT is dominated by the position uncertainties 
in sensor planes. 
In this case it is preferable to put all planes as close to each other as 
possible, and the {\bf N--N} configuration should be chosen.
In addition, there is an intermediate energy range
(depending on the assumed value of $\deldut$) where the {\bf WN--} 
configuration results in the best measurement.
In the high energy limit the multiple scattering can be neglected and
configurations {\bf W--W} and {\bf N--N} are expected to give 
same measurement precision.

\begin{figure}[p]
  \begin{center}
     $\Delta_{_{DUT}} = 800 \mu$m \\[-0.5cm]
     \epsfig{figure=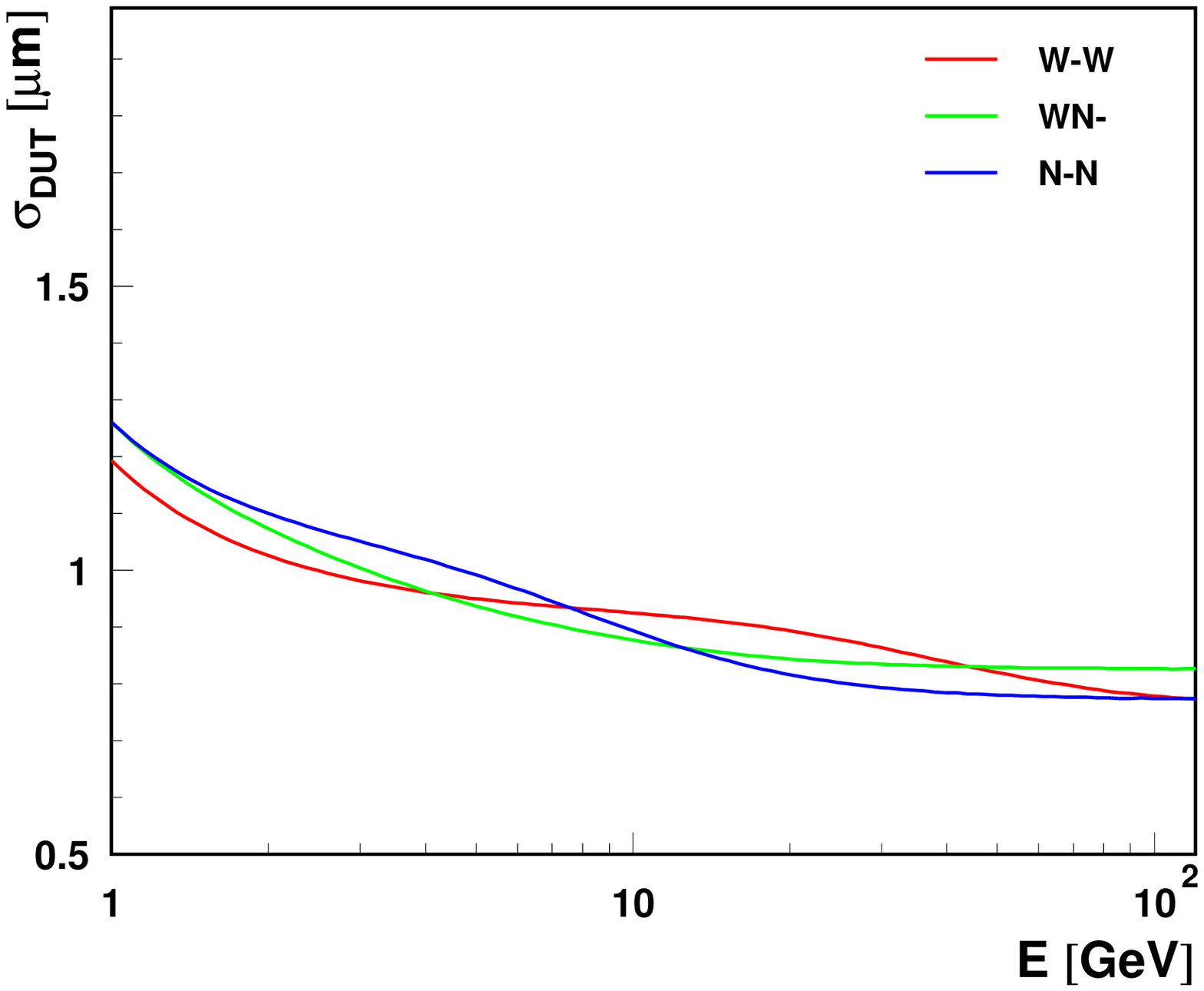,height=\figheight,clip=}\\
     $\Delta_{_{DUT}} = 300 \mu$m \\[-0.5cm]
     \epsfig{figure=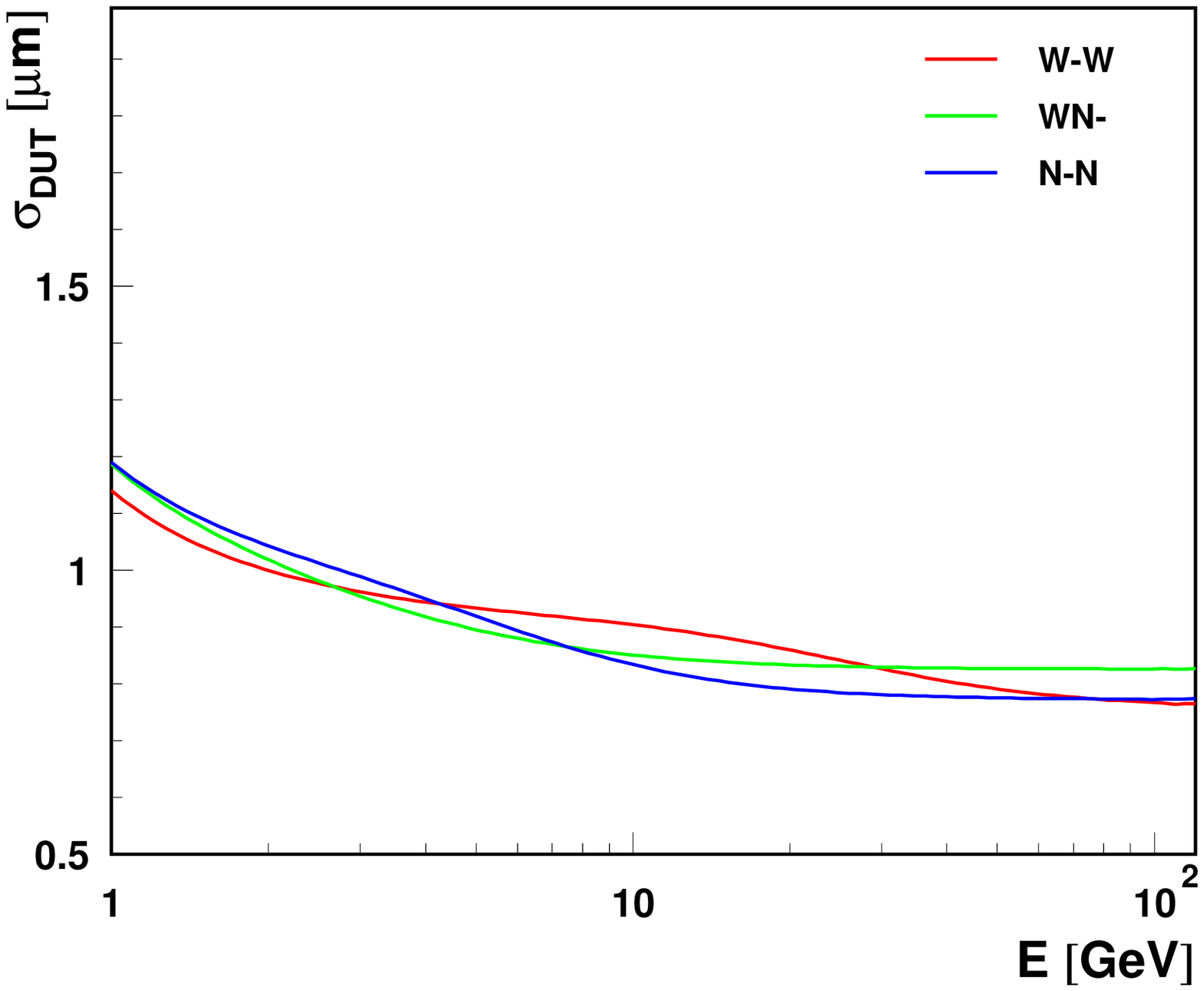,height=\figheight,clip=}
  \end{center}
 \caption{
Expected precision of position determination at DUT, $\sigdut$, 
as a function of the beam energy, 
for the assumed DUT thickness of 800$\mu m $ (upper plot) 
and 300$\mu m$ (lower plot) and the distance between 
the high-resolution plane and DUT of 2~mm.
Configurations with one high-resolution and three standard
sensor planes are considered. 
 }
 \label{fig:enecomp51_2} 
 \end{figure} 
If the distance between the high-resolution plane and DUT is
reduced to 2~mm, expected measurement precision improves,
as shown in \fig{enecomp51_2}.
Significant improvement is observed especially at low energies,
where large uncertainty is expected from multiple scattering.
Uncertainty in particle direction determination is large, bu
smaller distance reduces the error in extrapolating 
the position measured in the high-resolution plane to the
DUT surface.


\subsection{Results for 2+2 configuration}

\begin{figure}[tbp]
  \begin{center}
     \epsfig{figure=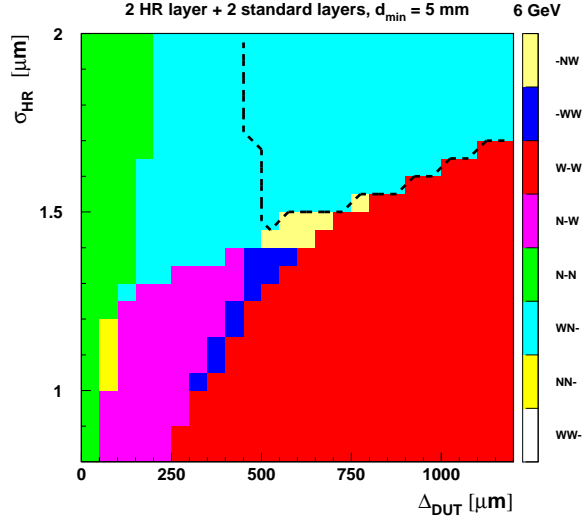,height=\figheight,clip=}
  \end{center}
 \caption{
Telescope configuration resulting in the is best position measurement
at DUT,  as a function of the assumed DUT thickness, $\deldut$ and the high-resolution plane measurement precision, $\sighr$, for electron beam
energy of 6~GeV.  Configurations with two high-resolution and two standard
sensor planes are considered. 
  }
 \label{fig:conf5_5_x_5} 
 \end{figure}

Analysis presented in the previous section was repeated for 
the telescope consisting of two  high-resolution planes 
and two standard sensor planes (2+2 configuration).
The optimal choice of telescope configuration for electron 
beam energy of 6~GeV, as a function of the assumed DUT 
thickness, $\deldut$, and the high-resolution plane measurement 
precision, $\sighr$, is shown in \fig{conf5_5_x_5}.
For $\sighr < 1.5 \mu m$ the high-resolution planes should
always be placed in front of and behind DUT, whereas for
large $\sighr$ and large $\deldut$ (the parameter range
indicated by dashed line in \fig{conf5_5_x_5}) best measurement
is obtained when both planes are put in front of DUT.

\begin{figure}[tbp]
\begin{center}
\begin{minipage}[t]{0.7\textwidth}
{\bf W--W}\\[-1cm]
\hspace*{2cm}
     \epsfig{figure=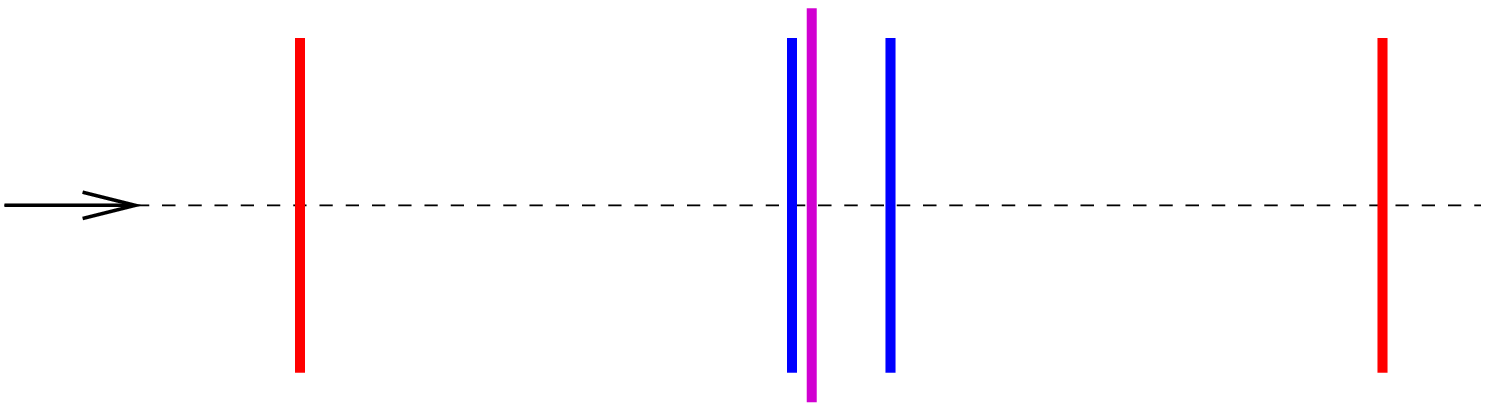,height=\diagsheight,clip=}\\[1cm]
{\bf N--W}\\[-1cm]
\hspace*{2cm}
     \epsfig{figure=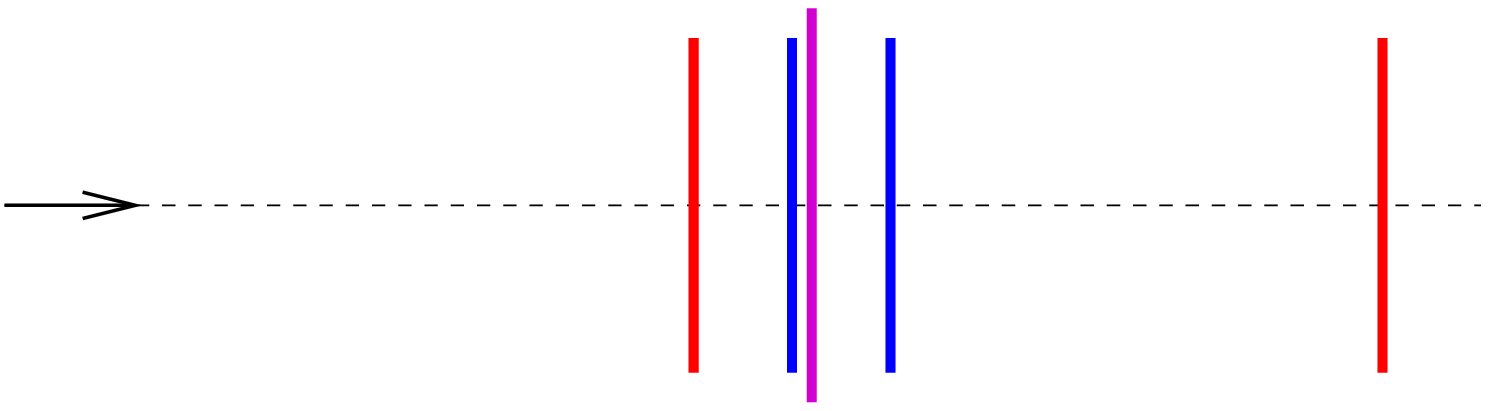,height=\diagsheight,clip=}\\[1cm]
{\bf NN--}\\[-1cm]
\hspace*{2cm}
     \epsfig{figure=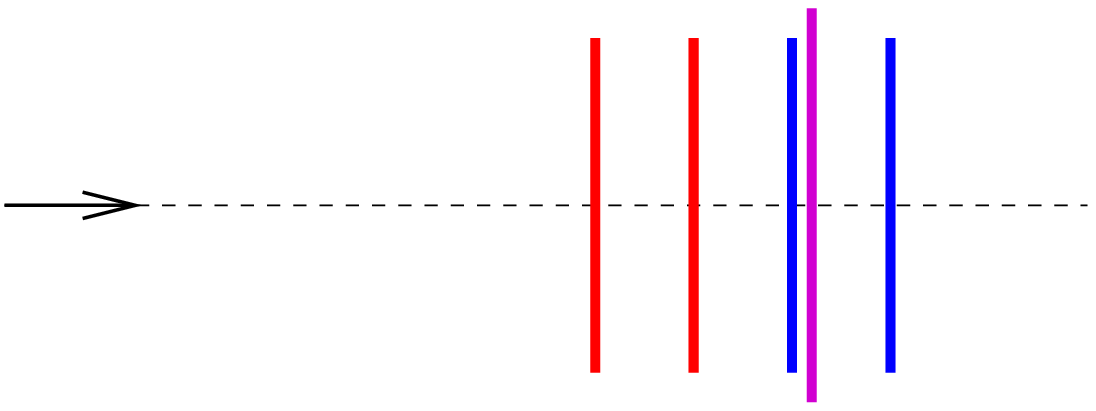,height=\diagsheight,clip=}\\[1cm]
{\bf N--N}\\[-1cm]
\hspace*{2cm}
     \epsfig{figure=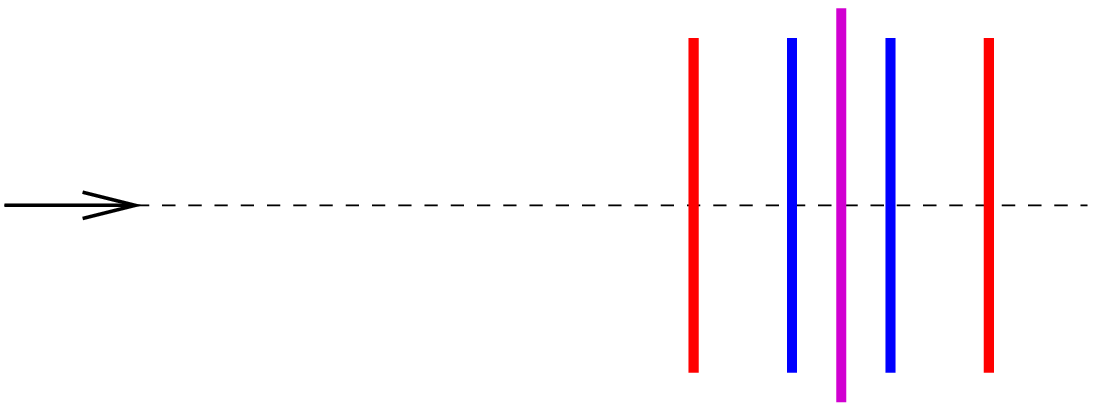,height=\diagsheight,clip=}
\end{minipage}
\end{center}
 \caption{
Configurations  resulting in the best position measurement
at DUT (in different ranges of parameters), 
for telescope consisting two high-resolution and two standard
sensor planes. High-resolution planes are indicated in blue, standard planes 
in red and the DUT in magenta.
 }
 \label{fig:conf_52} 
 \end{figure}

\begin{figure}[p]
  \begin{center}
     $\Delta_{_{DUT}} = 800 \mu$m \\[-0.5cm]
     \epsfig{figure=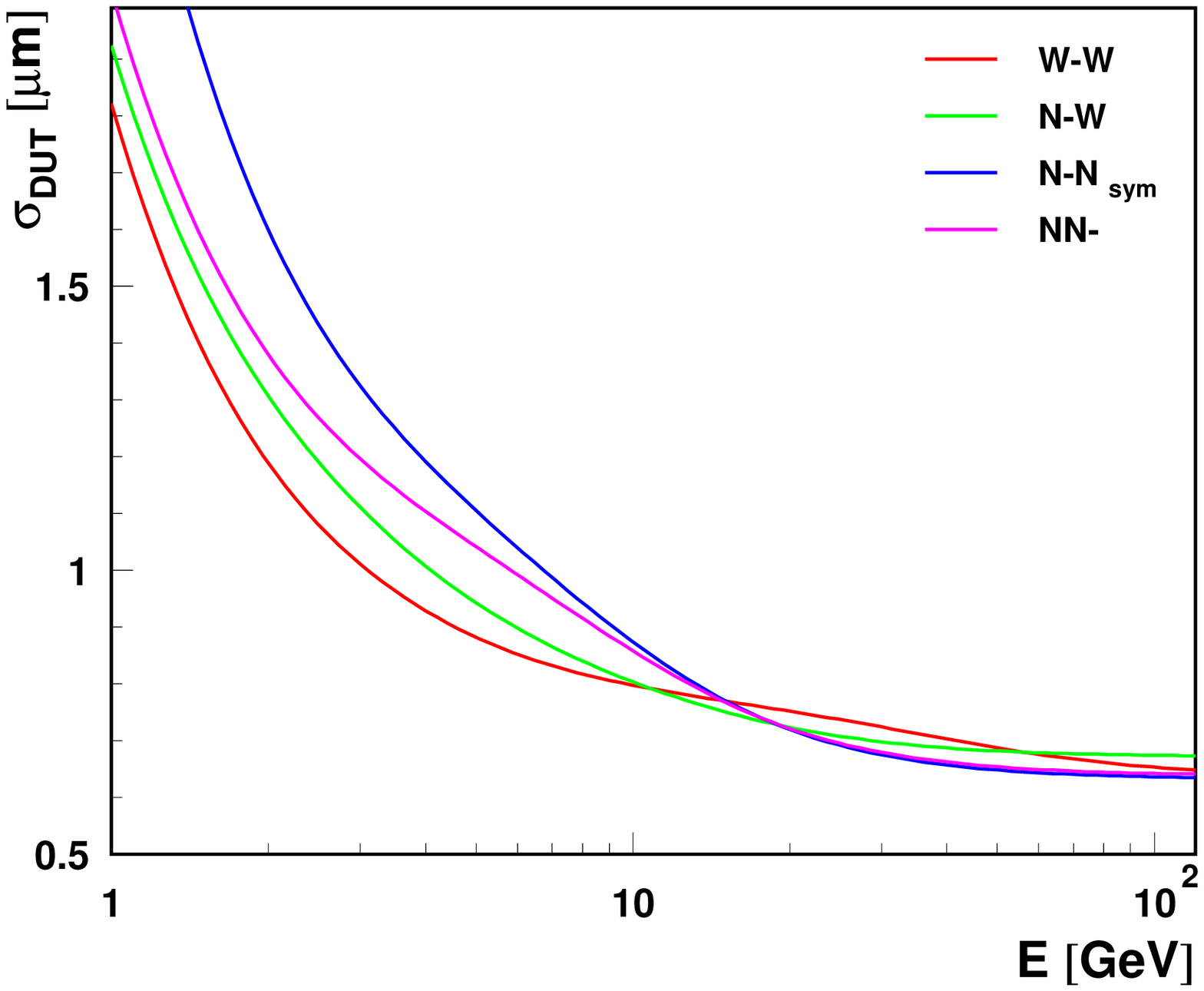,height=\figheight,clip=}\\
     $\Delta_{_{DUT}} = 300 \mu$m \\[-0.5cm]
     \epsfig{figure=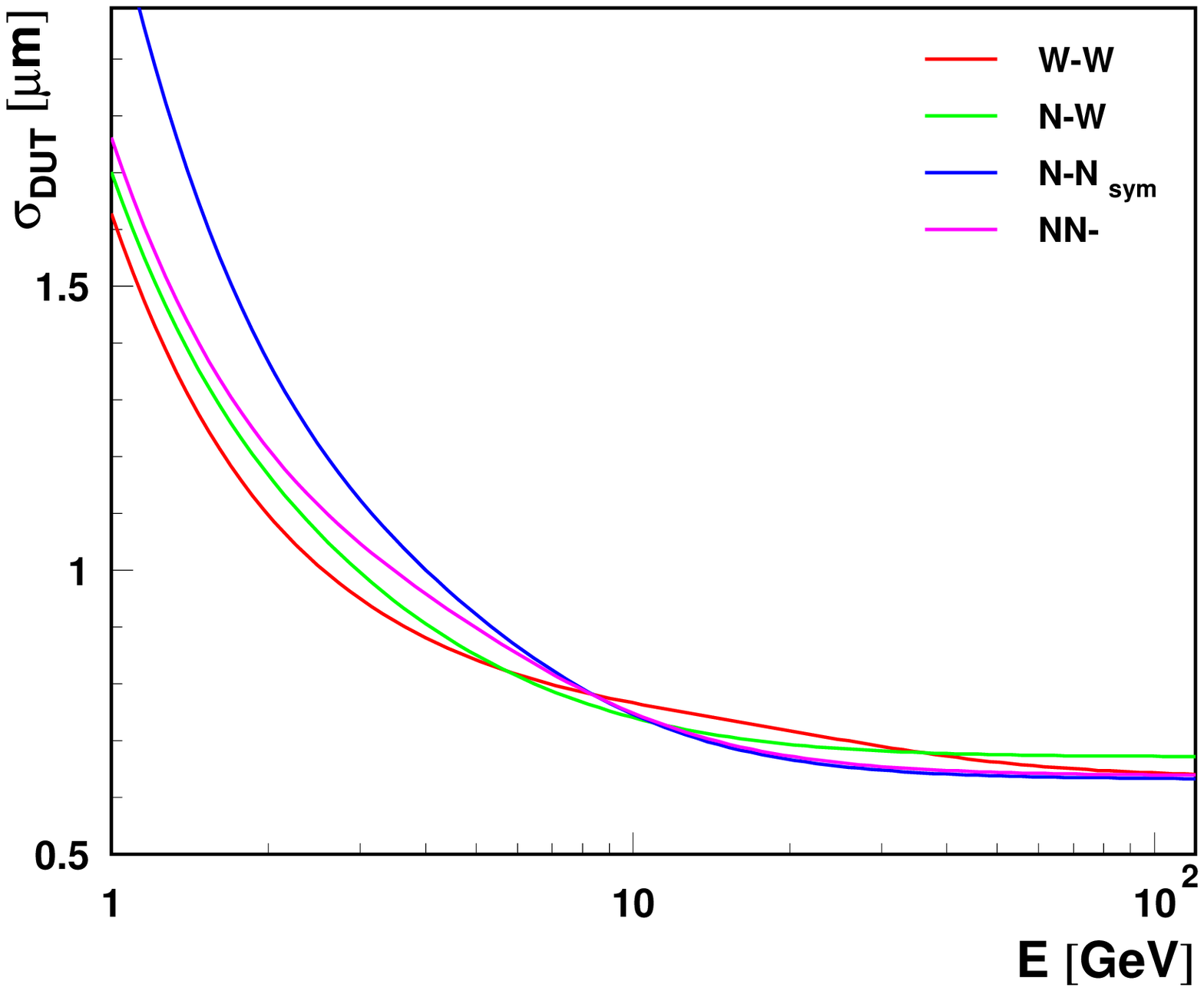,height=\figheight,clip=}
  \end{center}
 \caption{
Expected precision of position determination at DUT, $\sigdut$, 
as a function of the beam energy, 
for the assumed DUT thicknesses of 800$\mu m $ (upper plot) 
and 300$\mu m$ (lower plot) and the distance between 
the first high-resolution plane and DUT of 5~mm.
Configurations with two high-resolution and two standard
sensor planes are considered. 
 }
 \label{fig:enecomp52_5} 
 \end{figure}

Configurations  resulting in the best position measurement,
for different beam energies and thicknesses of DUT,
assuming high-resolution planes resolution $\sighr \sim 1 \mu m$
are shown in \fig{conf_52}.
The expected precision of position determination for these
telescope configuration, as a function of the beam energy, 
is shown in \fig{enecomp52_5}, for the DUT thicknesses 
of 800 and 300$\mu m$, and the distance between the first 
high-resolution plane and DUT of 5~mm.

The dependence of the optimal configuration choice of the beam energy
is similar to the case with one high-resolution plane.
At lowest energies, when the effects of multiple scattering are dominant,
best measurement is expected with {\bf W--W}  configuration.
At highest energies, when effects of the multiple scattering are small,
{\bf N--N} configuration should be chosen.
Contrary to the case with one high-resolution plane and 
to configurations optimal at lower energies,
in this case DUT should be placed in the middle between 
two high-resolution planes.

\begin{figure}[p]
  \begin{center}
     \epsfig{figure=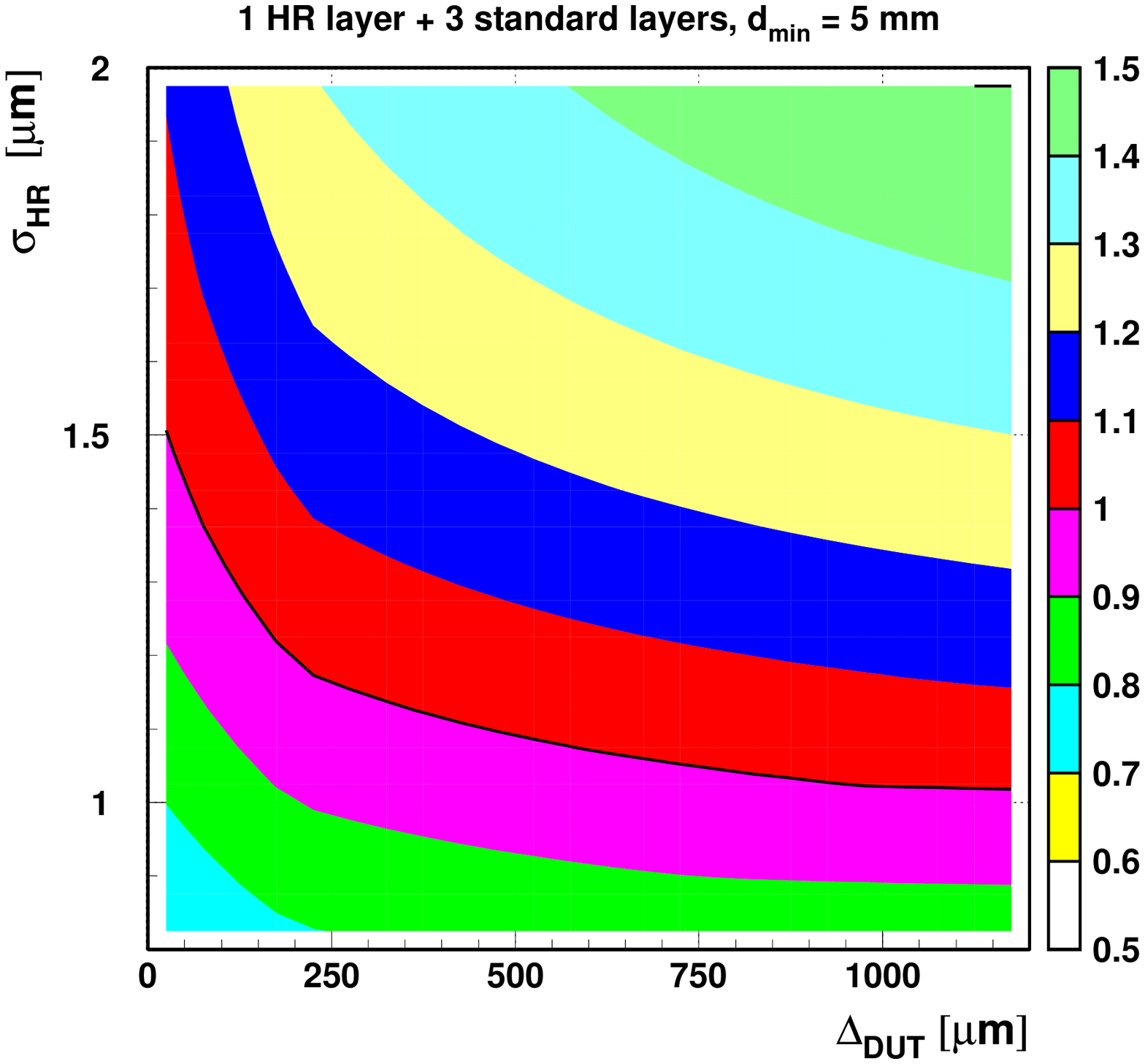,height=\figheight,clip=}\\
     \epsfig{figure=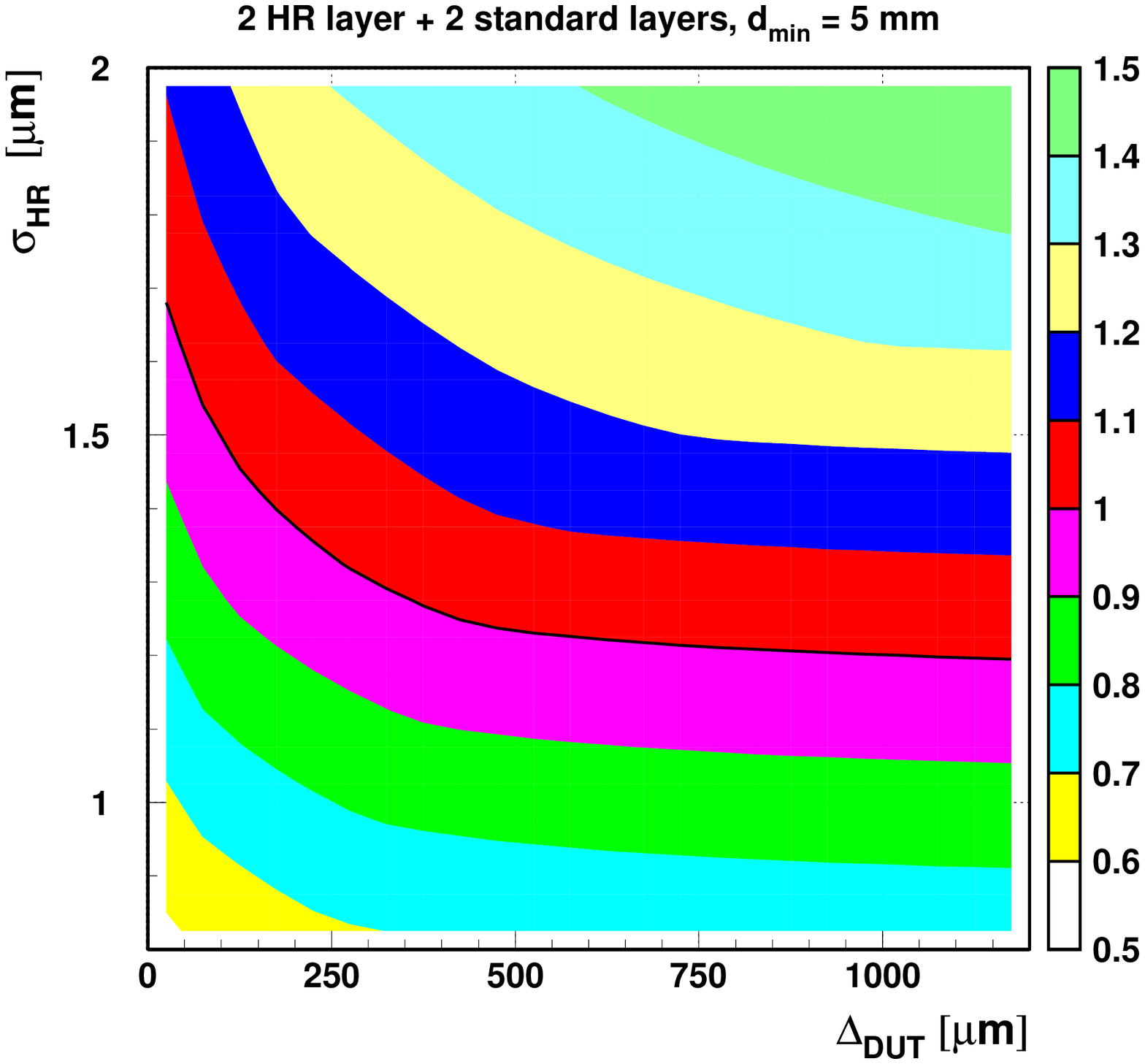,height=\figheight,clip=}
  \end{center}
 \caption{
Position resolution at DUT, $\sigdut$, obtained for
the optimal configuration choice, as a function of the assumed DUT 
thickness, $\deldut$ and the high-resolution plane measurement 
precision, $\sighr$. Configuration with one high-resolution 
and three standard sensor planes (upper plot) and with
two high-resolution and two standard sensor planes (lower plot)
are compared.
 }
 \label{fig:res5_2hr_2d} 
 \end{figure}

Including second high-resolution plane significantly improves position
resolution.
For beam energy of 6~GeV and high-resolution plane resolution 
$\sighr \sim 1 \mu m$ expected precision of position determination
improves by about 0.1$\mu m$. 
This is shown in \fig{res5_2hr_2d}
where the position resolution $\sigdut$ is shown as a function 
of the assumed DUT thickness, $\deldut$ and the high-resolution plane 
measurement precision, $\sighr$, as obtained in the configuration
optimal for given parameter values. 
Configurations with one and two high-resolution planes are compared.
Largest improvement due to second high-resolution plane
is expected at highest energies, when multiple scattering
can be neglected and the precision is determined by
the position measurement errors in telescope sensors
(compare \fig{enecomp51_5} and \fig{enecomp52_5}).

%
%

%
%

%
%

\clearpage


\subsection{Results for 1+5 and 2+4 configurations}

\begin{figure}[tbp]
 \begin{center}
\begin{minipage}[t]{0.7\textwidth}
{\bf NW--WN}\\[-1cm]
\hspace*{2cm}
    \epsfig{figure=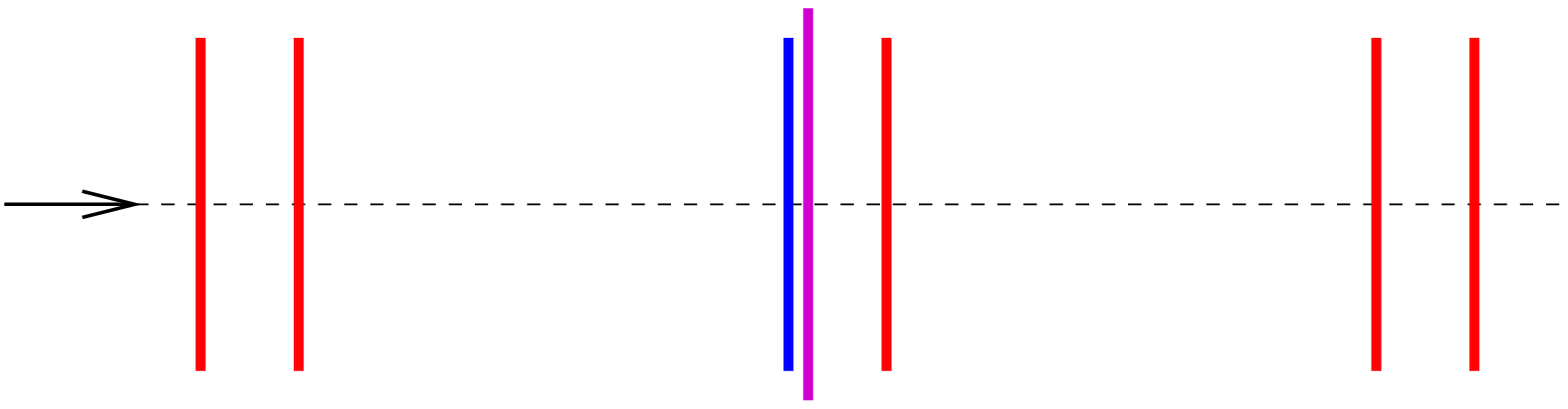,height=\diagsheight,clip=}\\[1cm]
{\bf WN--NW}\\[-1cm]
\hspace*{2cm}
     \epsfig{figure=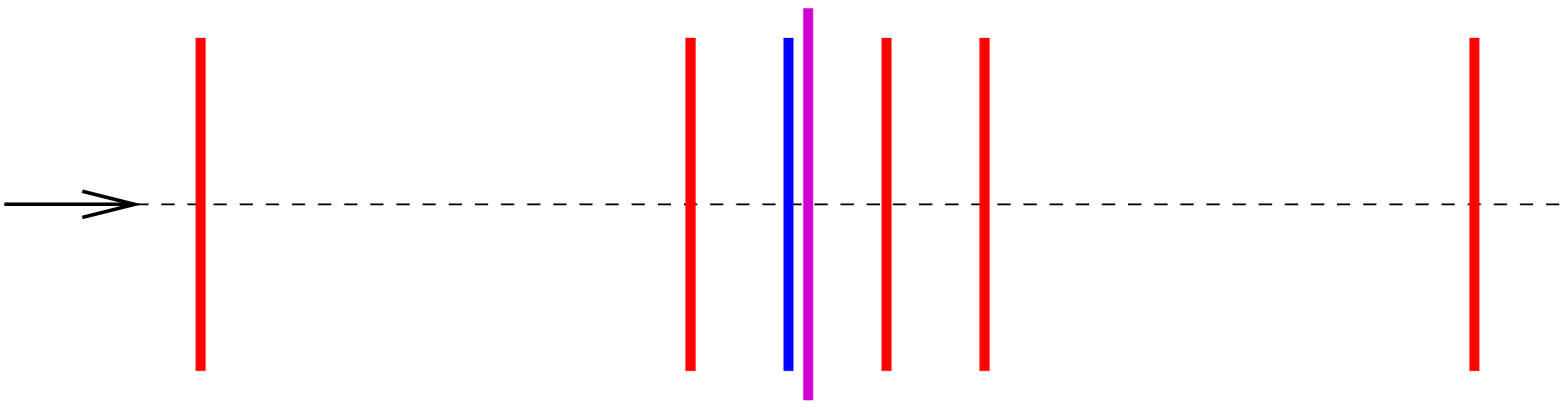,height=\diagsheight,clip=}\\[1cm]
{\bf WNN--N}\\[-1cm]
\hspace*{1.5cm}
     \epsfig{figure=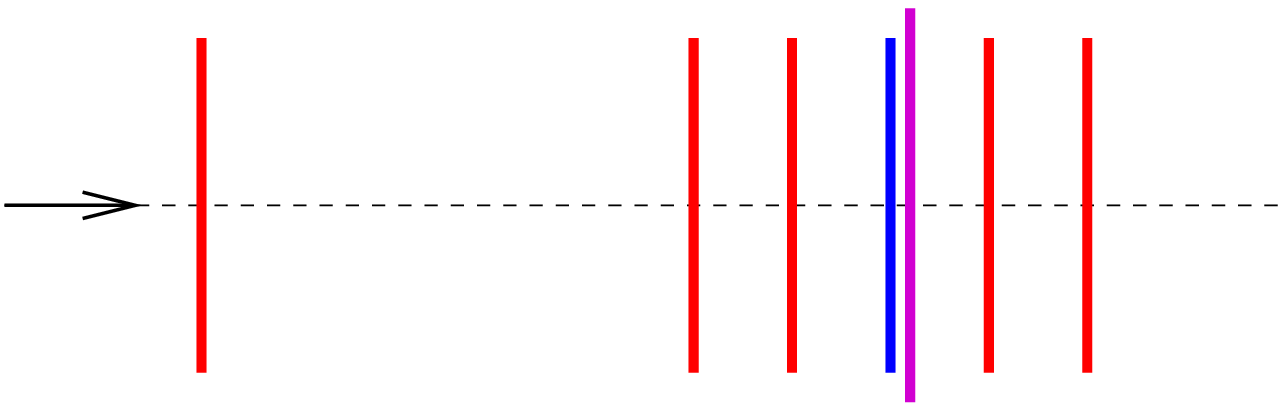,height=\diagsheight,clip=}\\[1cm]
{\bf NN--NN}\\[-1cm]
\hspace*{2.5cm}
     \epsfig{figure=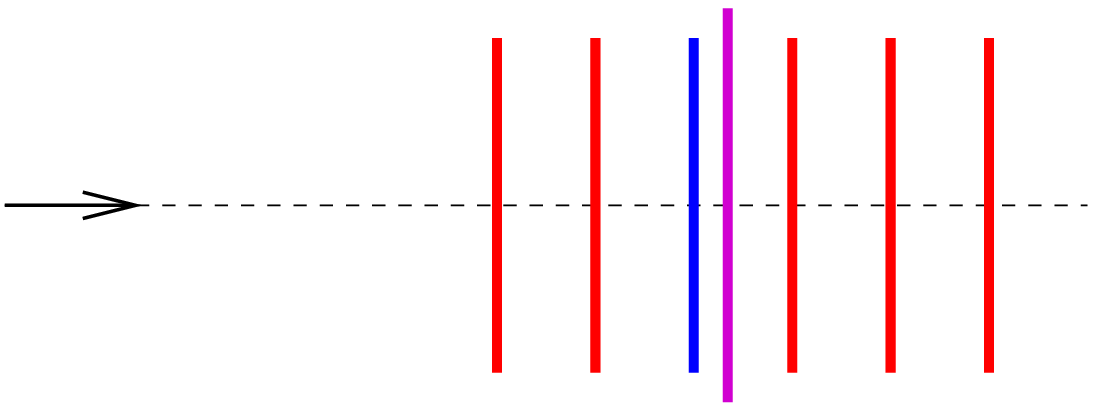,height=\diagsheight,clip=}
\end{minipage}
\end{center}
 \caption{
Configurations  resulting in the best position measurement
at DUT (in different ranges of parameters), 
for telescope consisting one high-resolution and five standard
sensor planes. High-resolution plane is indicated in blue, standard planes 
in red and the DUT in magenta.
 }
 \label{fig:conf_71} 
 \end{figure}

\begin{figure}[tbp]
 \begin{center}
\begin{minipage}[t]{0.7\textwidth}
{\bf NW--WN}\\[-1cm]
\hspace*{2cm}
     \epsfig{figure=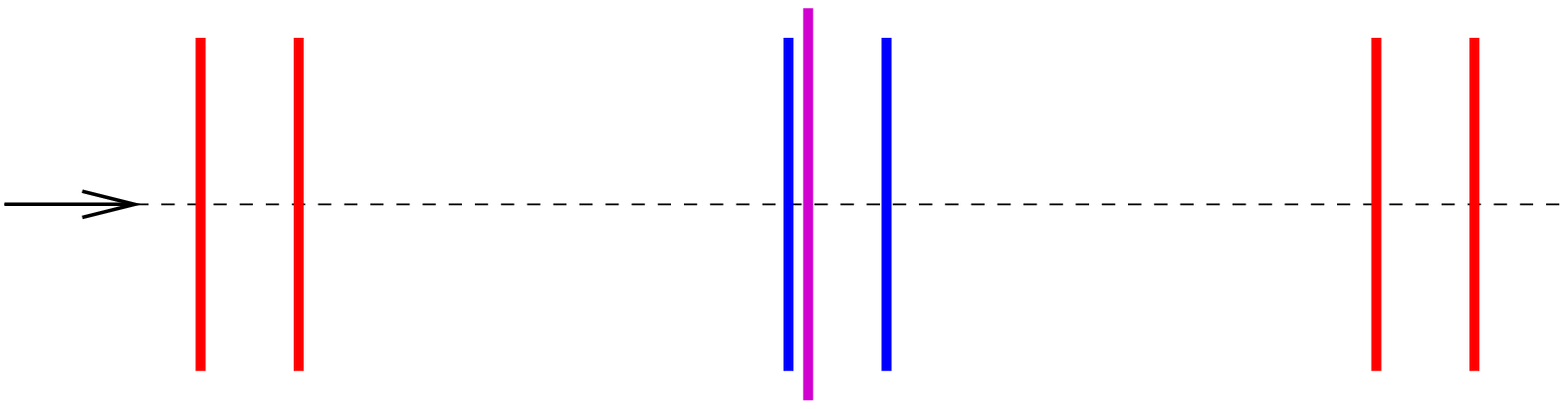,height=\diagsheight,clip=}\\[1cm]
{\bf WN--WW}\\[-1cm]
\hspace*{2cm}
     \epsfig{figure=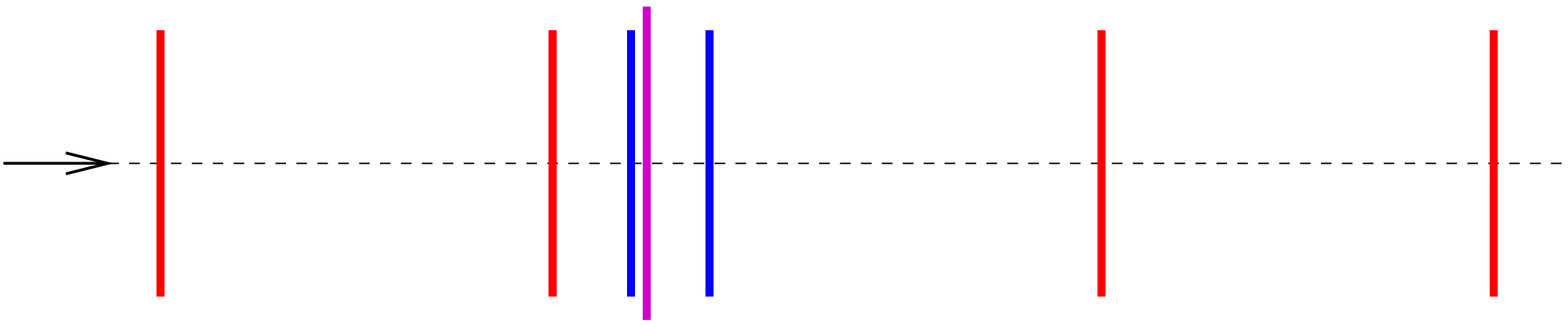,height=\diagsheight,clip=}\\[1cm]
{\bf N--NWW}\\[-1cm]
\hspace*{3.3cm}
     \epsfig{figure=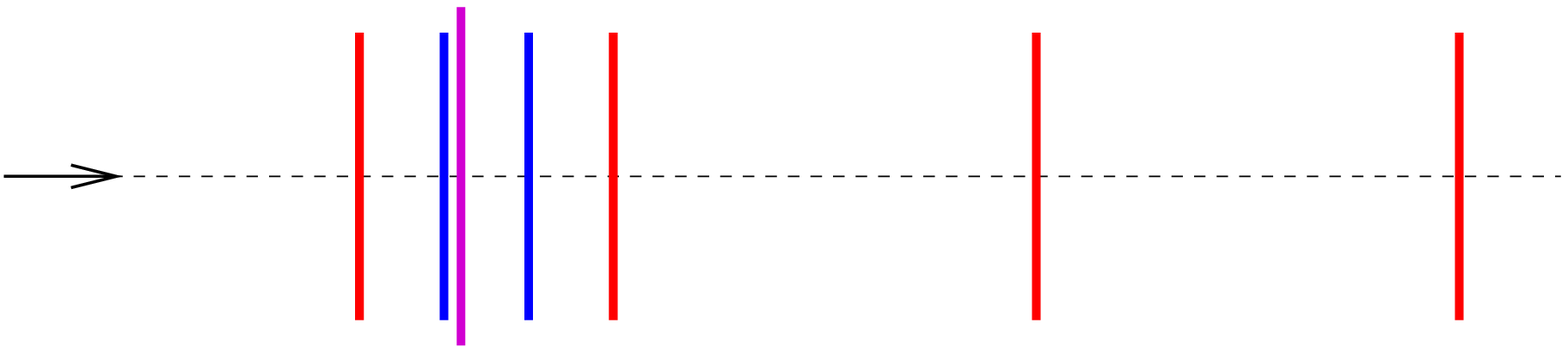,height=\diagsheight,clip=}\\[1cm]
{\bf NN--NN}\\[-1cm]
\hspace*{2.5cm}
     \epsfig{figure=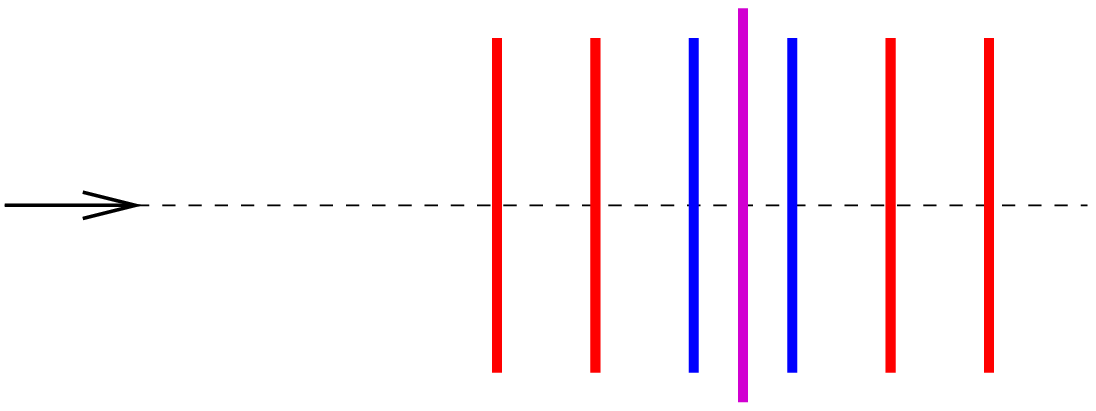,height=\diagsheight,clip=}
\end{minipage}
\end{center}
 \caption{
Configurations  resulting in the best position measurement
at DUT (in different ranges of parameters), 
for telescope consisting two high-resolution and four standard
sensor planes. High-resolution planes are indicated in blue, standard planes 
in red and the DUT in magenta.
 }
 \label{fig:conf_72} 
 \end{figure}

We also considered telescope consisting of 6 sensor planes
including one or two high resolution planes (configurations 
referred to as 1+5 and 2+4).
Configurations  resulting in the best position measurement,
for different beam energies and thicknesses of DUT,
assuming high-resolution planes resolution $\sighr \sim 1 \mu m$
are shown in \fig{conf_71} and \fig{conf_72} respectively.

%
%

\begin{figure}[p]
  \begin{center}
     $\Delta_{_{DUT}} = 800 \mu$m \\[-0.5cm]
     \epsfig{figure=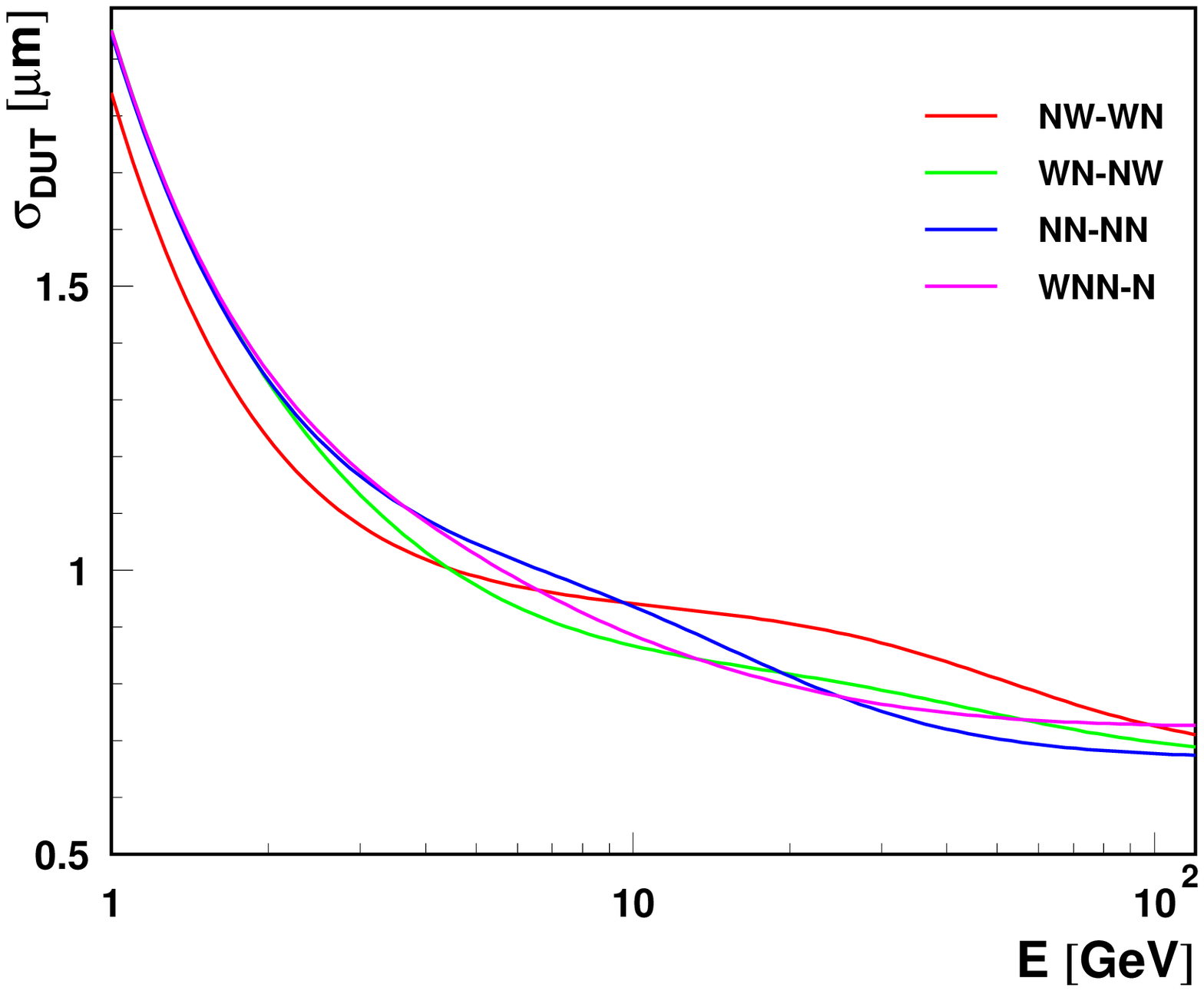,height=\figheight,clip=}\\
     $\Delta_{_{DUT}} = 300 \mu$m \\[-0.5cm]
     \epsfig{figure=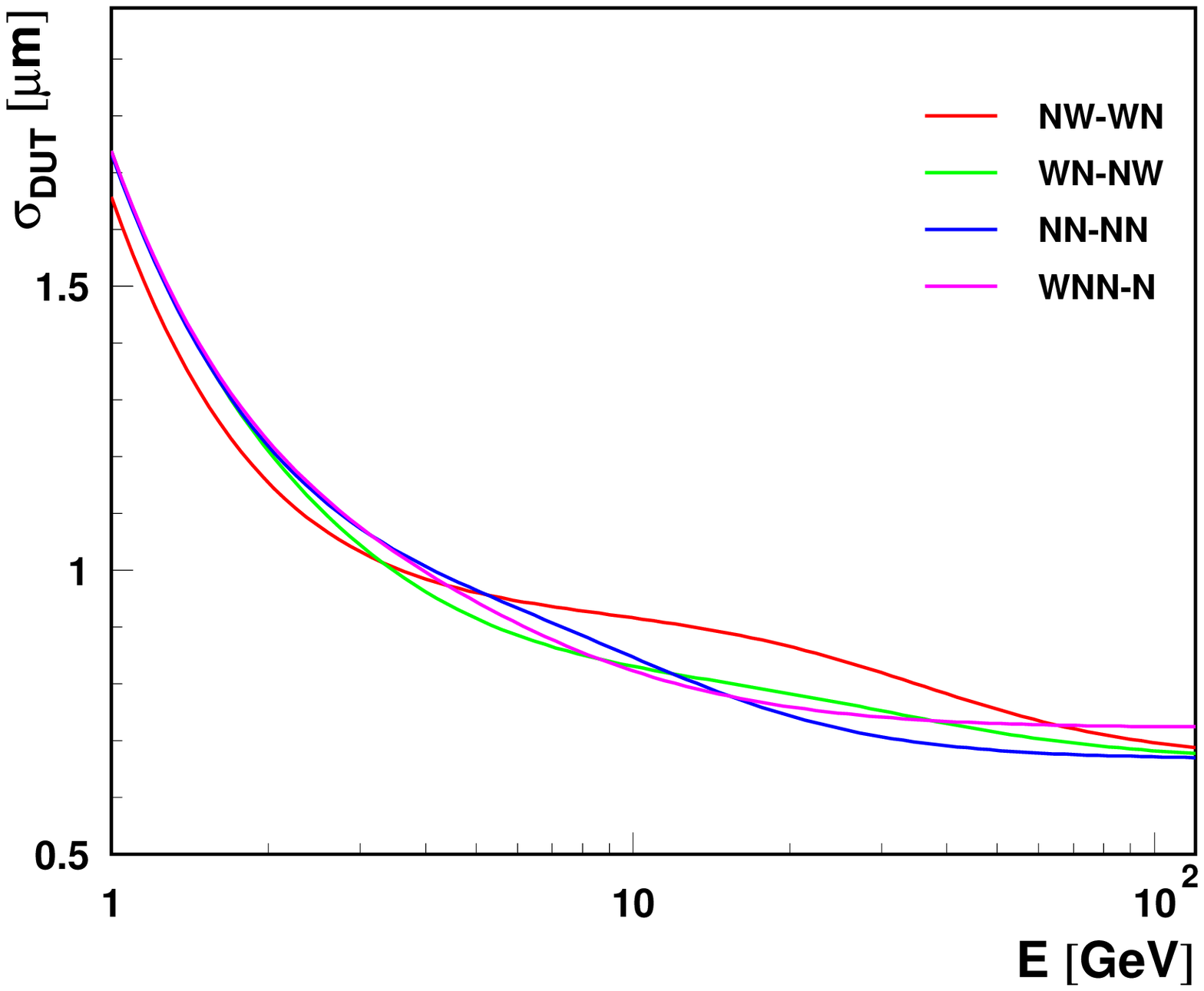,height=\figheight,clip=}
  \end{center}
 \caption{
 Expected precision of position determination at DUT, $\sigdut$, 
as a function of the beam energy, 
for the assumed DUT thicknesses of 800$\mu m $ (upper plot) 
and 300$\mu m$ (lower plot) and the distance between 
the high-resolution plane and DUT of 5~mm.
Configurations with one high-resolution and five standard
sensor planes are considered. 
 }
 \label{fig:enecomp71} 
 \end{figure}

\begin{figure}[p]
  \begin{center}
     $\Delta_{_{DUT}} = 800 \mu$m \\[-0.5cm]
     \epsfig{figure=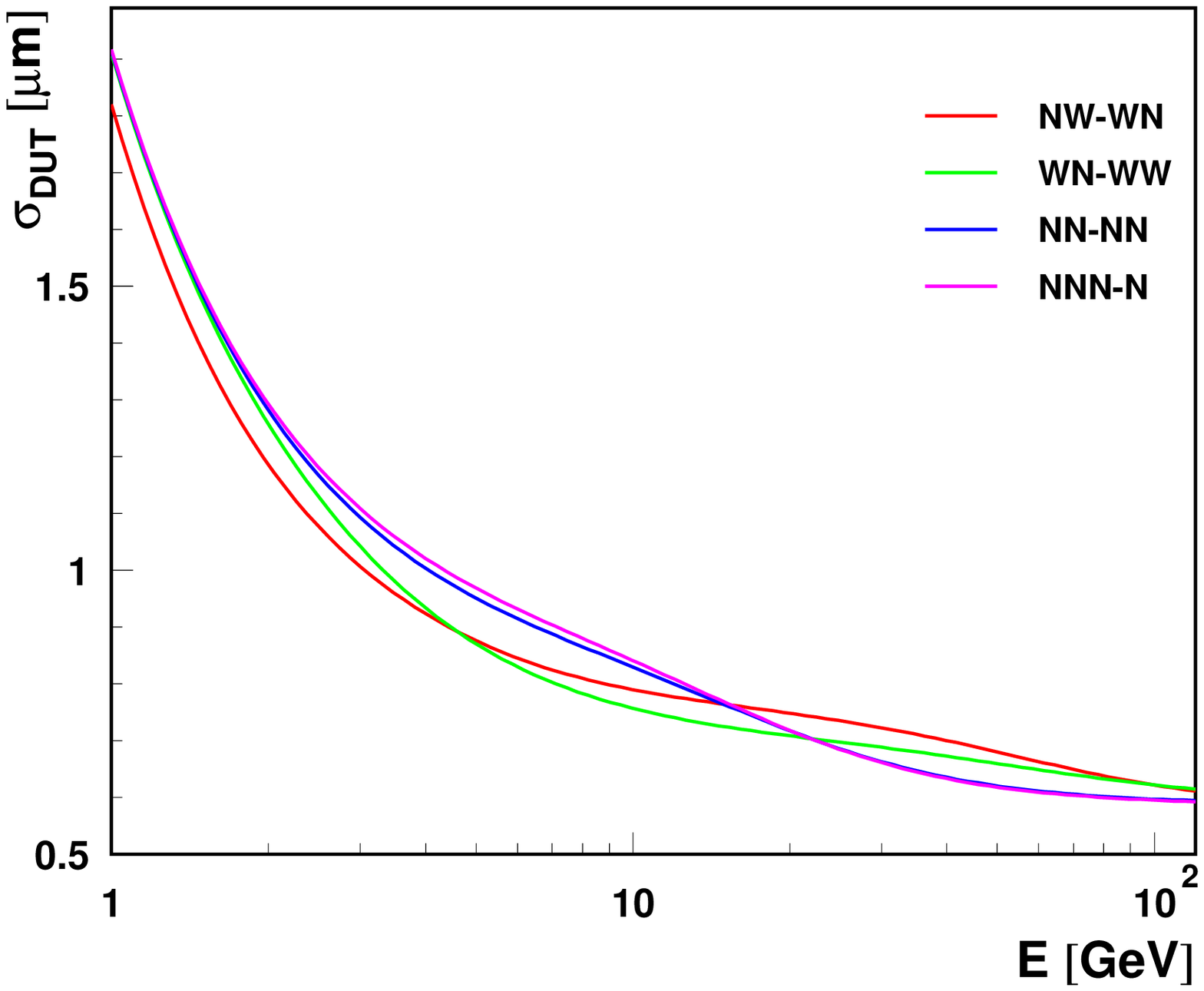,height=\figheight,clip=}\\
     $\Delta_{_{DUT}} = 300 \mu$m \\[-0.5cm]
     \epsfig{figure=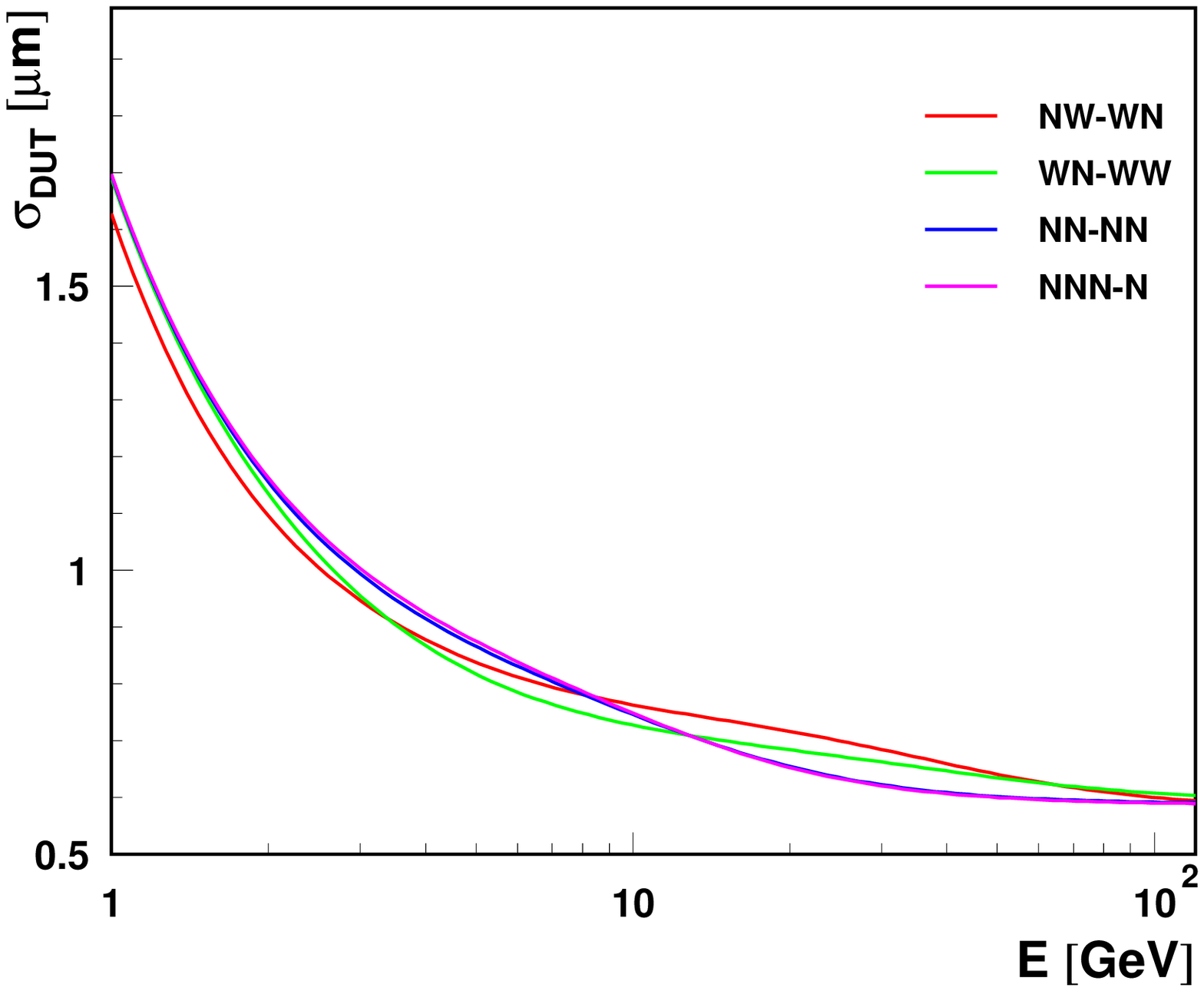,height=\figheight,clip=}
  \end{center}
 \caption{
Expected precision of position determination at DUT, $\sigdut$, 
as a function of the beam energy, 
for the assumed DUT thicknesses of 800$\mu m $ (upper plot) 
and 300$\mu m$ (lower plot) and the distance between 
the first high-resolution plane and DUT of 5~mm.
Configurations with two high-resolution and four standard
sensor planes are considered. 
 }
 \label{fig:enecomp72} 
 \end{figure}

The expected precision of position determination for 
these configurations, as a function of the beam energy 
is shown in \fig{enecomp71} and \fig{enecomp72}.
DUT thicknesses of 800 and 300$\mu m$ is considered and 
the distance between the (first) high-resolution plane 
and the DUT is 5~mm.

As for 1+3 and 2+2 configurations, the best measurement in the low
energy range is expected for the arrangement which provides long
"arms" for precise determination of particle directions before 
and after DUT.
This is required to minimize the error from multiple scattering on
the position extrapolation from the sensor planes to DUT.
For telescope consisting of six sensor layers it is obtained
for the {\bf NW--WN} configuration.
In the intermediate energy range three to five sensor planes
should be placed close to DUT.
The choice depends on the DUT thickness and differences between 
different configurations are not large.
In the high energy domain the situation is again similar to that
for the telescope with four sensor planes.
As the position error at DUT is dominated by the position uncertainties 
in sensor planes it is preferable to put all planes as close to each 
other as possible, which is denoted as {\bf NN--NN} configuration.
For the  {\bf NN--NN} configuration with two high-resolution planes (2+4) 
DUT should be placed in the middle between two high-resolution planes,
as indicated in \fig{conf_72}.

%
%

%
%

%
%

\subsection{Comparison of different telescope configurations}

It is clear that more telescope planes with better position resolution
always result in better precision for position determination at DUT.
However, it is important to understand which factor results in
a significant improvement of telescope performance and which gives only
a marginal effect.

\begin{figure}[p]
  \begin{center}
     \epsfig{figure=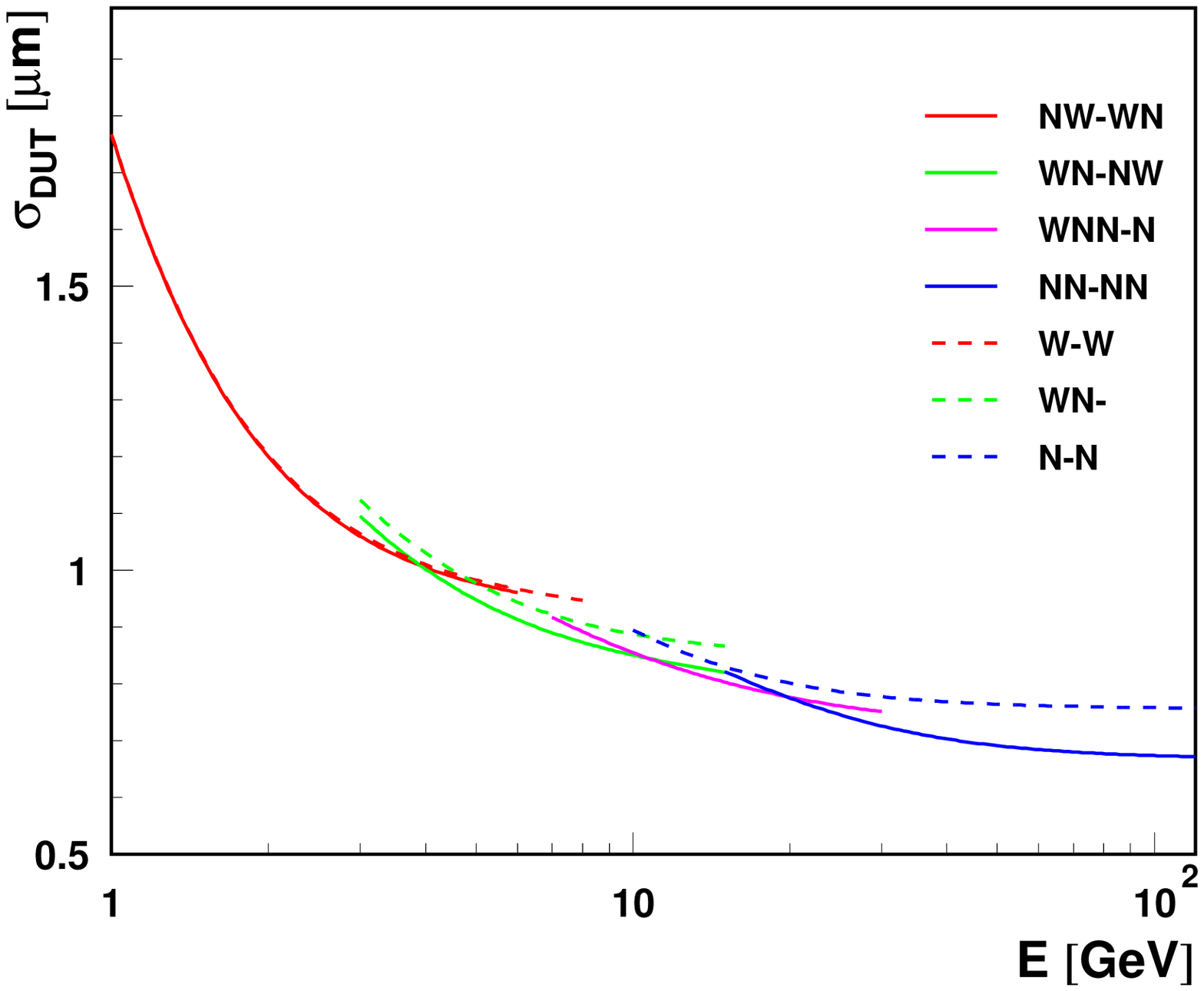,height=\figheight,clip=}\\
     \epsfig{figure=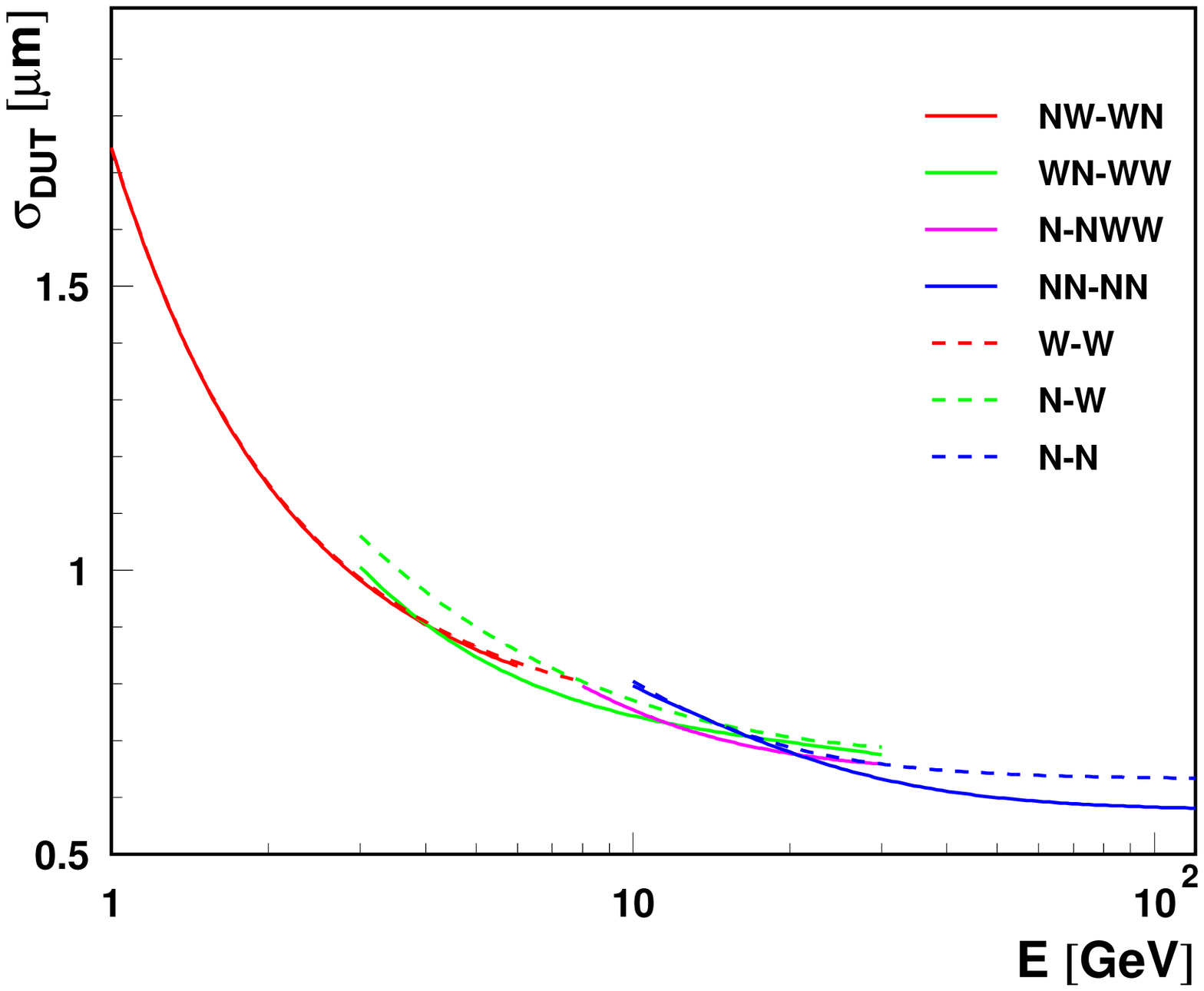,height=\figheight,clip=}
  \end{center}
 \caption{
Expected precision of position determination at DUT, $\sigdut$, 
as a function of the beam energy, 
for the assumed DUT thicknesses of 500$\mu m $ and the distance between 
the (first) high-resolution plane and DUT of 5~mm.
Configurations with six (solid lines) and four (dashed lines)
sensor planes are compared, for telescope with one (upper plot)
or two (lower plot) high-resolution planes.
 }
 \label{fig:enecomp752_5_500} 
 \end{figure}

Comparison of the expected position determination precision
for telescope with four and six sensor planes 
is shown in \fig{enecomp752_5_500}.
Two additional sensor planes improve position resolution at DUT
only for energies above 4~GeV, when more than two planes should
be placed close to DUT.
The improvement is largest at the highest energies.
For lowest energies, when "wide" configurations are preferred
only marginal improvement is observed between {\bf W--W} and
{\bf NW--WN} configurations.
%
%

\begin{figure}[p]
  \begin{center}
     \epsfig{figure=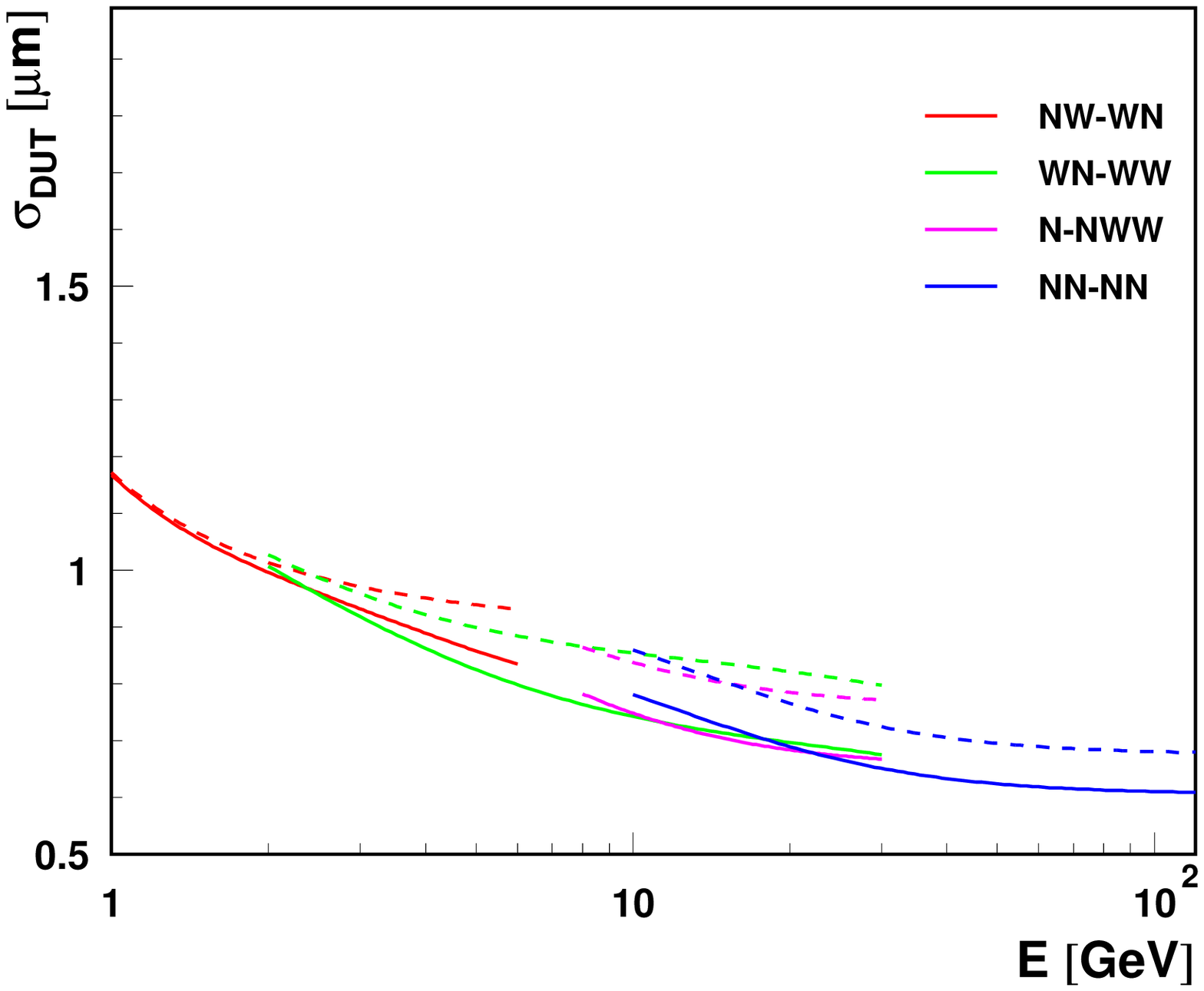,height=\figheight,clip=}\\
     \epsfig{figure=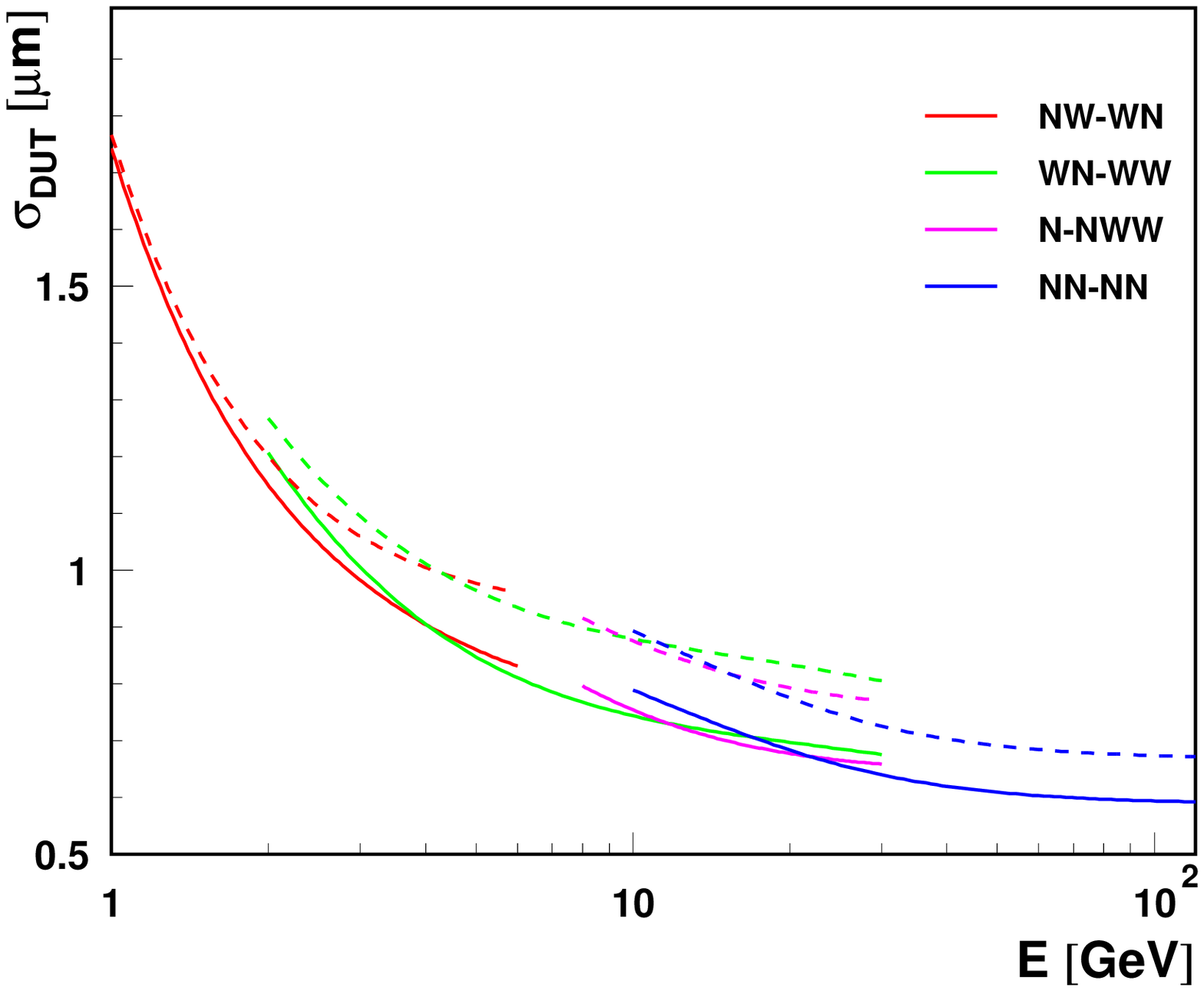,height=\figheight,clip=}
  \end{center}
 \caption{
Expected precision of position determination at DUT, $\sigdut$, 
as a function of the beam energy, 
for the assumed DUT thicknesses of 500$\mu m $ and the distance between 
the (first) high-resolution plane and DUT of 2~mm (upper plot) 
and 5~mm (lower plot).
Configurations with two (solid lines) and one (dashed lines)
high-resolution planes are compared, for telescope consisting
of six sensor planes.
 }
 \label{fig:enecomp721_5_500} 
 \end{figure}

Large improvement, up to about  0.15$\mu m$, is expected due to second 
high-resolution plane, as shown in \fig{enecomp721_5_500}.
Both at the intermediate and at high beam energies 
precision obtained with two high-resolution sensors is
significantly better than with one high-resolution sensor only.
In the intermediate energy range, the effect is stronger
for the larger distance between DUT and first high-resolution plane.
As before, for lowest beam energies, below about 2~GeV,
improvement due to second high-resolution plane is marginal.
In this energy range the precision is determined by the measurement
in the closest telescope plane.

\begin{figure}[p]
  \begin{center}
     \epsfig{figure=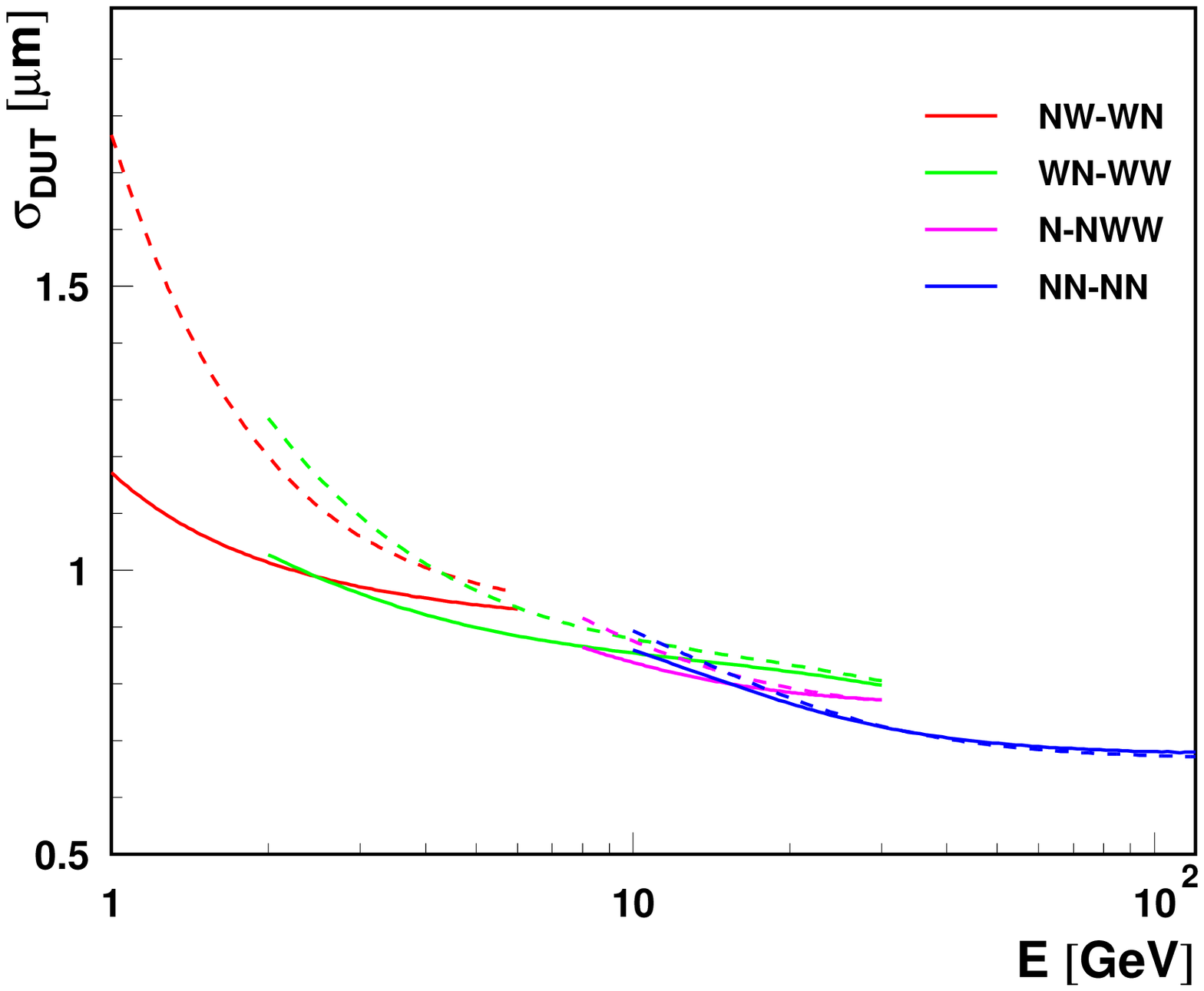,height=\figheight,clip=}\\
     \epsfig{figure=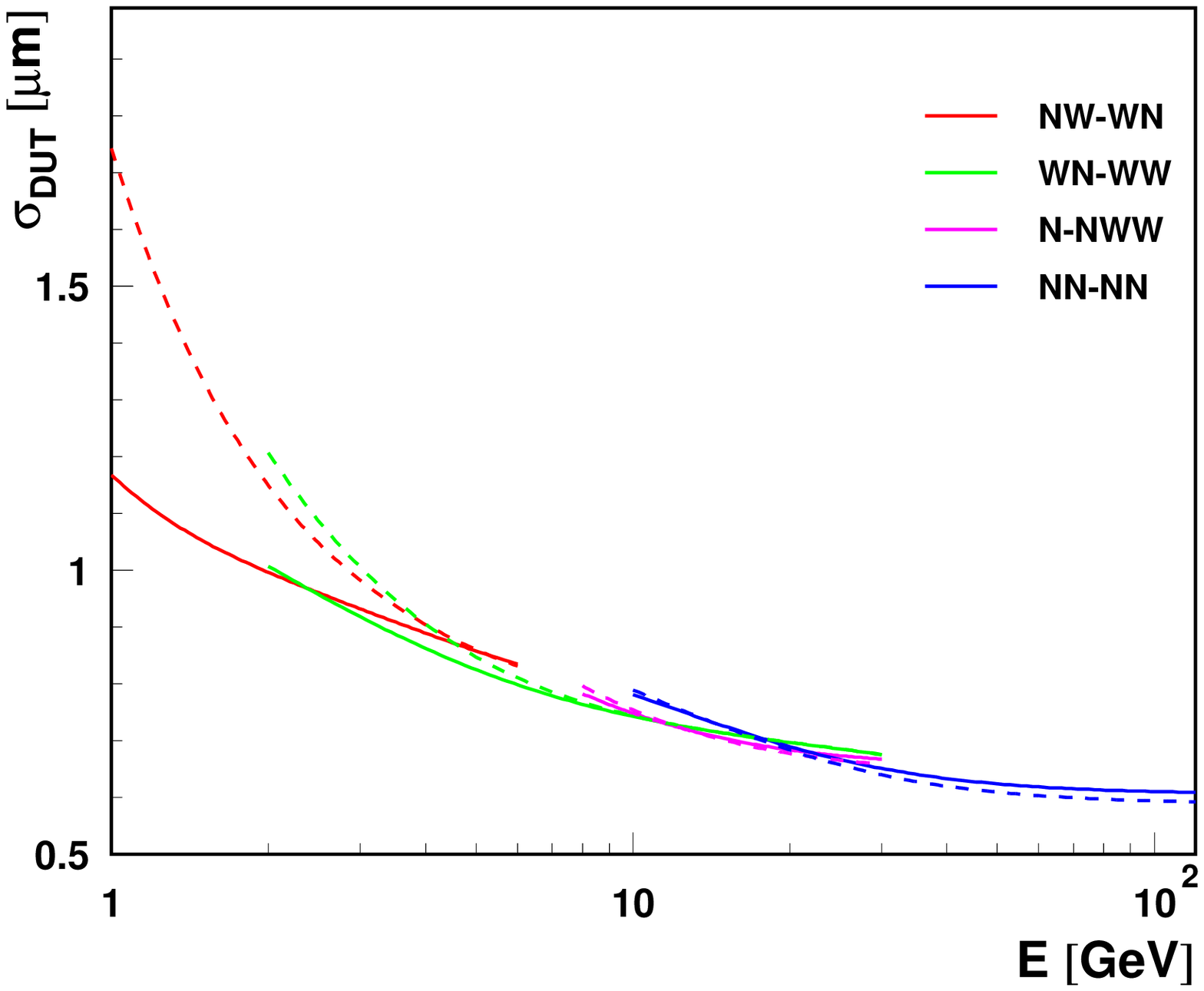,height=\figheight,clip=}
  \end{center}
 \caption{
Expected precision of position determination at DUT, $\sigdut$, 
as a function of the beam energy, 
for the assumed DUT thicknesses of 500$\mu m $.
Configurations with the assumed distance between 
the (first) high-resolution plane and DUT of 2~mm (solid line) 
and 5~mm (dashed line) are compared, for telescope 
consisting of six sensor planes 
including one (upper plot) or two (lower plot)
high-resolution planes.
 }
 \label{fig:enecomp7_2_5_500} 
 \end{figure}

Show in \fig{enecomp7_2_5_500} is a comparison of expected position
resolution for different distances between the high-resolution plane
and DUT.
For lowest beam energies minimizing this distance is crucial
for the precise position determination.
By reducing the distance between first high-resolution plane and DUT
from 5 to 2 mm the position uncertainty can be reduced by over 30\%.
With one high-resolution plane improvement is observed up to energies
of about 20~GeV.
For highest beam energies configuration with 2~mm distance gives results
slightly worse than with 5~mm.
This is because more symmetric plane setup is preferred when multiple 
scattering can be neglected.
This is event more visible for configurations with two high-resolution planes.
For highest energies best position determination is obtained when DUT
is placed in the middle between two high-resolution planes.

\begin{figure}[p]
  \begin{center}
     \epsfig{figure=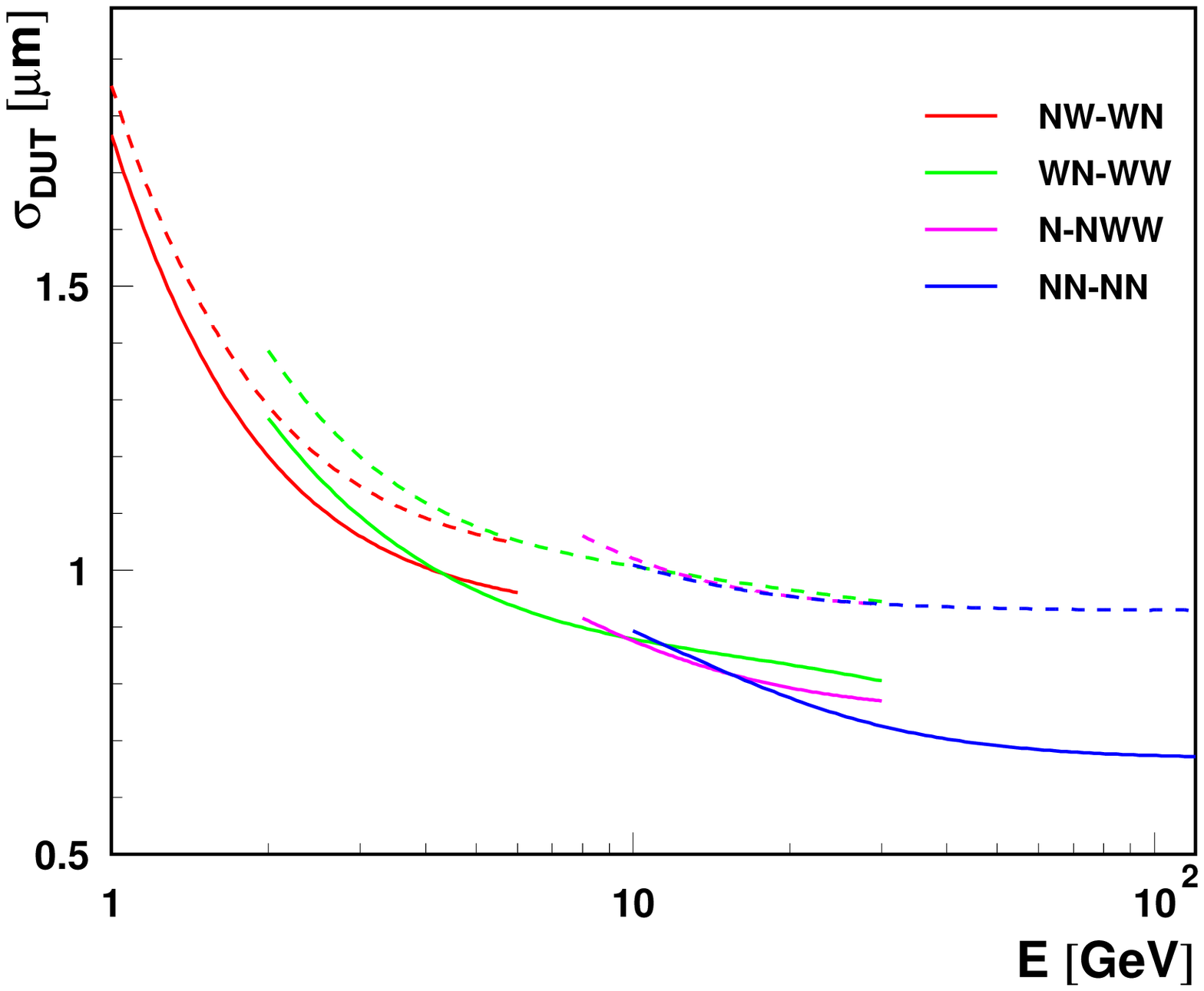,height=\figheight,clip=}\\
     \epsfig{figure=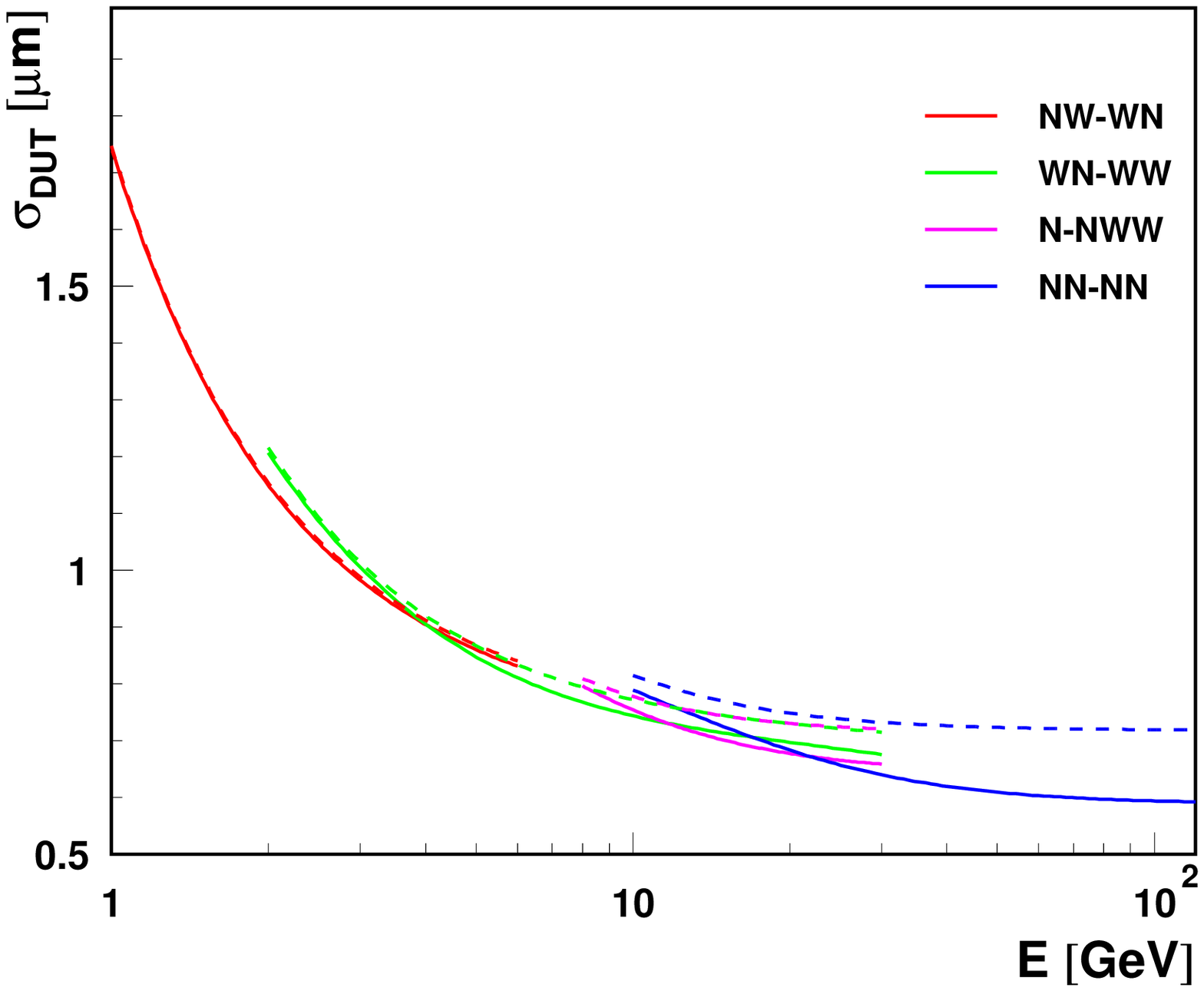,height=\figheight,clip=}
  \end{center}
 \caption{
Expected precision of position determination at DUT, $\sigdut$, 
as a function of the beam energy, 
for the assumed DUT thicknesses of 500$\mu m $.
Configurations with the assumed standard plane position resolution
of  2~$\mu m$ (solid line) and 5~$\mu m$ (dashed line) are compared, 
for telescope consisting of six sensor planes 
including one (upper plot) or two (lower plot)
high-resolution planes.
Distance between the (first) high-resolution plane and DUT is 5~mm.
 }
 \label{fig:tel7hr} 
 \end{figure}

One of the considered solutions for standard telescope sensors
are the sensors with binary readout. 
Proposed $16 \mu m$ pitch corresponds to the position resolution
of about 5$\mu m$ (see also Section \ref{sec:geantcomp}).
Expected precision of position determination at DUT,
for standard planes with 2~$\mu m$ and 5~$\mu m$ resolution,
is compared in \fig{tel7hr}. 
For telescope with one high-resolution plane, resolution of
standard planes have large impact on the expected position
reconstruction precision.
Effects of up to 30\% are expected at high energies.
However, if telescope is equipped with two high-resolution planes,
no significant loss of resolution is expected below 5~GeV.


\section{Telescope alignment studies}

All results presented so far were obtained assuming perfect telescope
alignment.
It could be expected that they remain valid only if the sensor
alignment is performed with precision much better than the single
plane resolution.
Surprisingly, proposed track fitting method turned out to be little
sensitive to telescope misalignment.

Simulation results described in Section~\ref{sec:sim} has been used
to verify influence of telescope alignment on the track fitting.
Many 'experiments' were performed by generating random sensor shifts 
(in the plane perpendicular to the beam direction) according to the
assumed alignment accuracy.
For each set of plane positions simulated sensor responses were
reevaluated and track fit was repeated for every {\sc Geant~4} event.
Longitudinal position uncertainties should have much smaller effect
on the telescope performance and were not considered.
Possible plane rotations were also not considered in the presented study.

\begin{figure}[tbp]
  \begin{center}
     \epsfig{figure=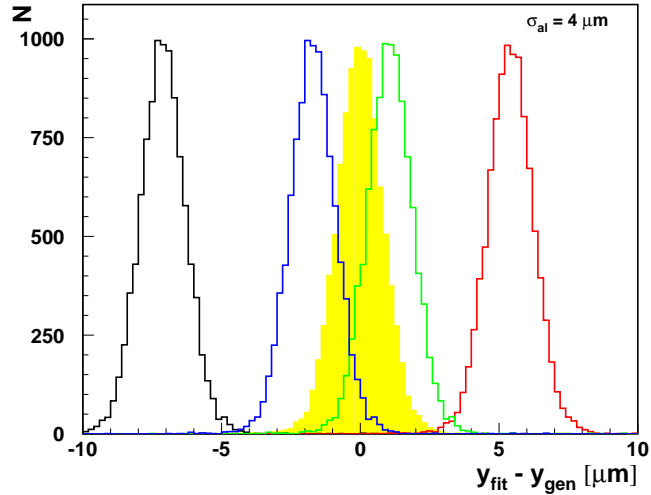,height=\figheight,clip=}
  \end{center}
 \caption{
Difference between reconstructed and true particle position at DUT, for
6~GeV electron beam. Open histograms show simulation results for four
experiments with the assumed alignment uncertainty of 4$\mu m$.
For comparison, results obtained assuming perfect telescope alignment are
shown as a filled histogram.
 }
 \label{fig:al_shift} 
 \end{figure}

Shown in \fig{al_shift} are the obtained distributions of position
reconstruction error at DUT (the difference between reconstructed 
and true particle position), for four different sets of telescope 
plane positions generated assuming alignment uncertainty of 3$\mu m$.
For comparison, results obtained assuming perfect telescope alignment are
also shown.
%
%
The telescope misalignment results only in the systematic shift in the
position determined at DUT.
This shift is equivalent to the shift in DUT position.
The width of the distribution, which determines the precision of 
the measurement is unchanged.
This follows from the equation \eqref{fit}.
Reconstructed position at DUT is given by a linear combination of 
positions measured in telescope layers, with coefficients (elements
of matrix {\cal S}) depending on the telescope geometry and position
resolutions in single planes.
Therefor, any constant shift in measured position results in a constant
shift in the reconstructed position at DUT.

\begin{figure}[tbp]
  \begin{center}
     \epsfig{figure=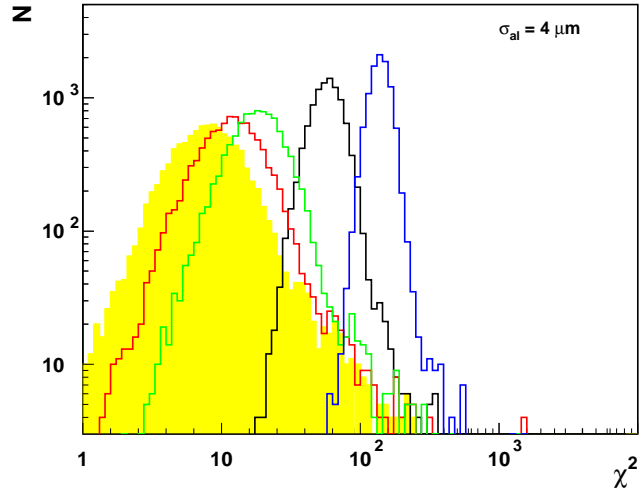,height=\figheight,clip=}
  \end{center}
 \caption{
$\chi^2$ distribution for simulated {\sc Geant 4} events, for
6~GeV electron beam. Open histograms show results for four
experiments with the assumed alignment uncertainty of 4$\mu m$.
For comparison, results obtained assuming perfect telescope alignment are
shown as a filled histogram.
 }
 \label{fig:al_chi2} 
 \end{figure}

Although the telescope resolution is not directly affected,
alignment uncertainties do influence the track reconstruction.
This is because plane misalignment results in large  $\chi^2$ values,
not related to the actual measurement.
This is illustrated in \fig{al_chi2}, where $\chi^2$ distributions
for simulated {\sc Geant 4} events are shown for  four
experiments with the assumed alignment uncertainty of 4$\mu m$.
For comparison, results obtained assuming perfect telescope alignment are
also shown.
%
%
In two of this experiments no event would pass the $\chi^2 < 20$ cut
which was imposed to select well fitted events when perfect
alignment was assumed (see Section \ref{sec:sim}).
We have to conclude that with large alignment uncertainties
$\chi^2$ cut can no longer be used to remove 
poorly reconstructed tracks. 
Therefor, effective telescope resolution can deteriorate slightly.
It will also be much more difficult to match hits to the track,
if multiple hits are reconstructed in single telescope layer.
In order to be able to use $\chi^2$ as a measure of track quality
we should reduce alignment error to the level comparable with 
position resolution in single telescope planes i.e. few $\mu m$.

\begin{figure}[tbp]
  \begin{center}
     \epsfig{figure=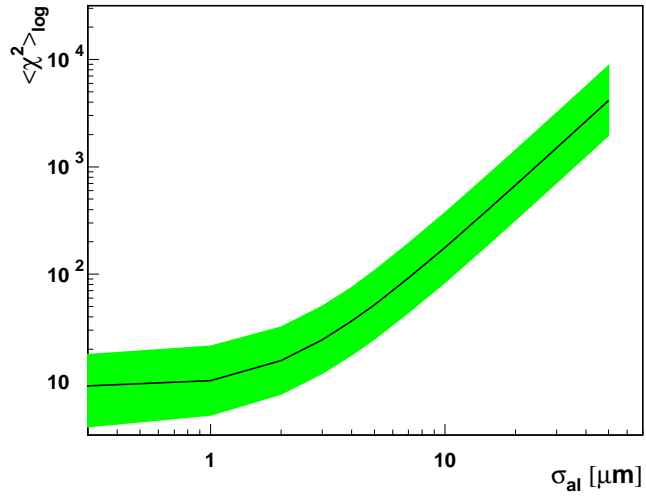,height=\figheight,clip=}
  \end{center}
 \caption{
$\chi^2$ value corresponding to the value of $\log \chi^2$
averaged over many Monte Carlo experiments,
as a function of the assumed  alignment uncertainty, $\sigma_{al}$.
Band width corresponds to the RMS of  $\log \chi^2$ distribution.
 }
 \label{fig:al_meanchi} 
 \end{figure} 

Precise telescope alignment, taking into account not only transverse  
but also longitudinal sensor shifts and rotations, requires dedicated
analysis of collected data.
However, based on the described track fitting algorithm we can propose 
a very simple method, which can be used to verify the transverse
plane alignment with adequate accuracy.

The expected beam spot size is too large to allow for precise plane
positioning. 
Also the beam direction can not be used as a constraint, due to 
angular spread of incident beam and multiple scattering in subsequent planes
(especially at low energies).
Therefor only relative plane positioning is possible.
We have to choose two planes and use them as the reference
for aligning the remaining sensors.

\begin{figure}[tbp]
  \begin{center}
     \epsfig{figure=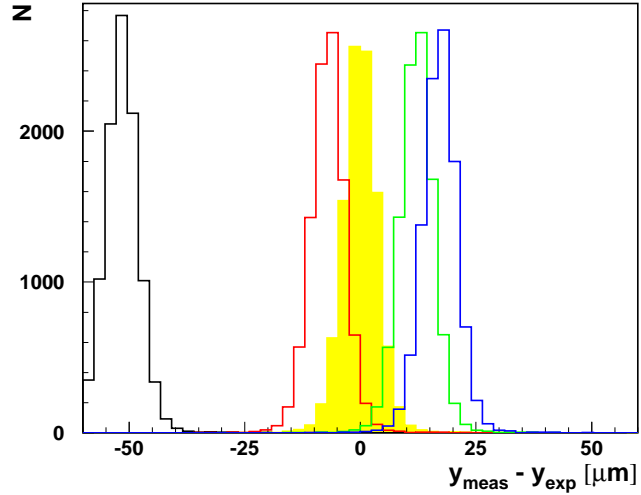,height=\figheight,clip=}
  \end{center}
 \caption{
Difference between particle position measured in the 3rd telescope plane
and the expected position, as obtained from the fit to the second 
and the last but one plane, for 6~GeV electron beam. 
Open histograms show simulation results for four
experiments with the assumed alignment uncertainty of 10$\mu m$.
For comparison, results obtained assuming perfect telescope alignment are
shown as a filled histogram.
 }
 \label{fig:al_hist2} 
 \end{figure}

For telescope consisting of seven planes (DUT and six sensors) 
the best choice is to use second and the last but one plane as
a reference.
Only for the highest beam energies, when the multiple scattering
can be neglected, selecting first and last plane gives slightly better results.
Results of multiple {\sc Geant 4} simulations 
with the assumed 10 $\mu m$ alignment 
uncertainty are shown in \fig{al_hist2}.
For each experiment, difference between the particle position measured 
in the third telescope plane and the expected position (from the fit
to the second and last but one plane) has an approximately Gaussian 
distribution with a width of about 3.6$\mu m$.
From this distribution position of the plane can be established
with accuracy of the order of $0.1 \mu m$ already with about
1000 reconstructed particle tracks.

\begin{figure}[tbp]
  \begin{center}
     \epsfig{figure=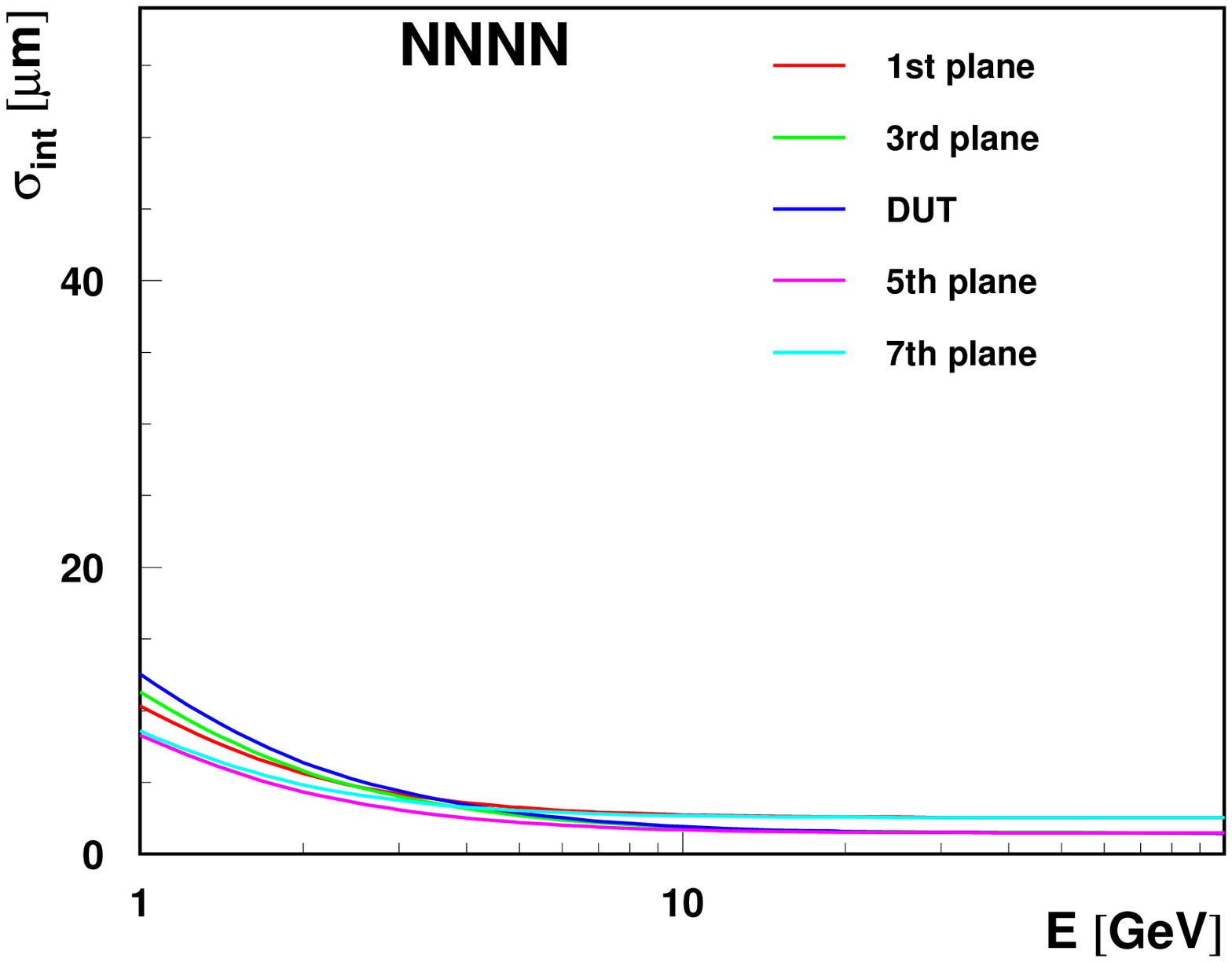,height=\figheight,clip=}
     \epsfig{figure=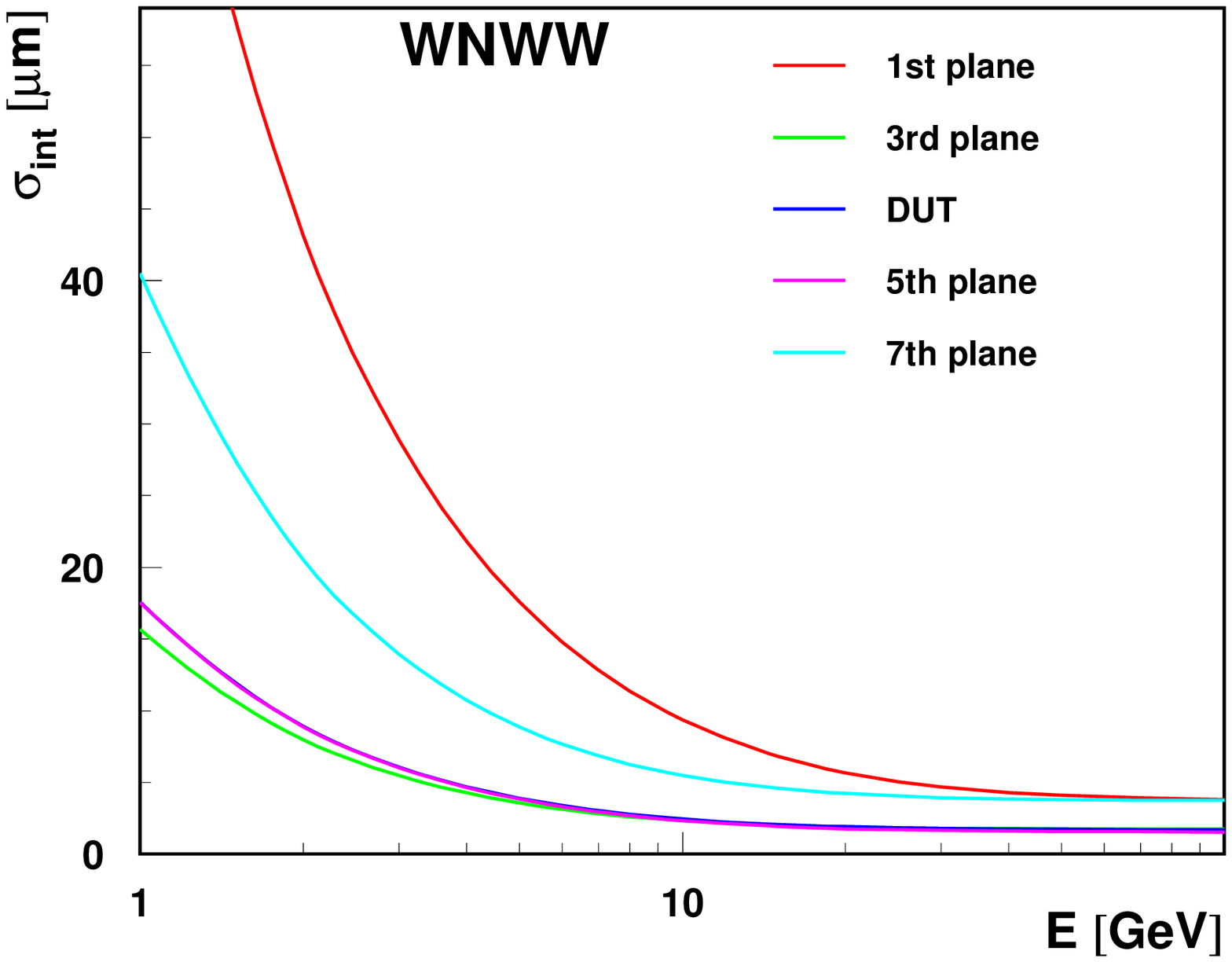,height=\figheight,clip=}
  \end{center}
 \caption{
Expected precision of particle position determination from
line fit to measurements in the second and the last but one telescope planes
as a function of beam energy, for the assumed DUT thicknesses of 500$\mu m $.
Position determination in remaining telescope planes as well as in DUT
are considered for {\bf NN-NN} (upper plot)  and {\bf WN-WW} (lower plot)
configurations.
 }
 \label{fig:al_dtest} 
 \end{figure}

Shown in \fig{al_dtest} is the expected precision of particle 
position determination from the considered fit to two planes,
as a function of beam energy. 
Two telescope configurations are considered, corresponding to
smallest ({\bf NN-NN}) and largest ({\bf WN-WW}) fit errors.
Even for wide telescope configuration and low beam energies,
plane alignment with accuracy of the order of 1$\mu m$ should
be possible with reasonable event statistics.
It is clear that the alignment procedure gives best results
for highest beam energy available.

\section{Conclusion}

The main aim of the presented study was to  
understand the position measurement in the EUDET pixel telescope and
suggest the optimum telescope plane setup, 
resulting in the best position measurement.
Analytical method for particle track fitting 
with multiple scattering has been developed 
and  verified using {\sc Geant 4} simulation.
The method gives qualitative improvement as compared to straight line fits, 
and allows to use all collected events in the analysis analysis.
The approach allows for analytical calculation
of the expected measurement uncertainty, 
so the telescope performance could be studied in detail
without time-consuming MC simulation.

Telescope performance was studied for setups consisting
of four or six readout planes, including one or two high-resolution sensors.
For each setup considered, different telescope configurations
(corresponding to different distances between planes
or different plane order) has been compared.
It turned out that the optimum plane configuration is not uniquely defined
and depends not only on the telescope parameters but also on 
the beam energy and the assumed DUT thickness.
However, differences between configurations optimal at different
parameter ranges are not large.
If one configuration has to be chosen, configuration with wide
telescope ``arms'' ({\bf W--W} or {\bf NW--WN}) should be used, 
as it gives much larger gain in resolution at low energies 
than the loss at higher energies (as compared to other configurations).

It was also confirmed that with appropriate track fitting method
configuration with 6 sensor planes 
always give better position resolution than 4 planes, 
but the difference is significant only at high energies.

One of the important advantages of the analytical 
track fitting method is that it is hardly sensitive to 
telescope misalignment.
Transverse displacements of telescope sensors result
only in the systematic shift in the reconstructed particle position
at DUT, but the position resolution remains unchanged. 
Therefor, possible small  telescope misalignment affects 
only the track quality estimate based on $\chi^2$ calculation.
For proper selection of good tracks plane alignment 
with accuracy of few $\mu m$ is sufficient.
This can be achieved with a simple procedure of relative plane alignment,
based on track fitting to two selected planes.

\section*{Acknowledgement}
This work is partially supported 
by the Polish  Ministry of Science and Higher Education, 
project number EUDET/217/2006 (2006-2009)
and by the Commission of the European Communities
under the 6$^{th}$ Framework Programme "Structuring the European
Research Area", contract number RII3-026126.

\end{document}